\documentclass[twocolumn,trackchanges]{aastex62} %linenumbers

\usepackage{threeparttable} 
\graphicspath{{./}{figures/}}

\received{???}
\revised{???}
\accepted{???}
\shorttitle{Surveying the Milky Way Substructures and Streams}
\shortauthors{Wang et al.}

\begin{document}
\title{Galactic-Seismology Substructures and Streams Hunter with LAMOST and Gaia. I. \\ Methodology and Local Halo Results}
\author{Guan-Yu Wang}
\affil{Department of Astronomy, China West Normal University, Nanchong, 637002, P.\,R.\,China}
\author[0000-0001-8459-1036]{Hai-Feng Wang}
\affil{Department of Astronomy, China West Normal University, Nanchong, 637002, P.\,R.\,China}
\affil{Centro Ricerche Enrico Fermi, Via Panisperna 89a, I-00184 Rome, Italy}
\affil{Department of Physics, Sapienza, Universit\'a di Roma, P.le A. Moro 5, I-00185 Rome, Italy}
\affil{GEPI, Observatoire de Paris, Universit\'e PSL, CNRS, Place Jules Janssen F-92195, Meudon, France}
\author{Yang-Ping Luo}
\affil{Department of Astronomy, China West Normal University, Nanchong, 637002, P.\,R.\,China}
\author{Yuan-Sen Ting}
\affil{Research School of Astronomy \& Astrophysics, Australian National University, Canberra, ACT 2611, Australia}
\affil{School of Computing, Australian National University, Acton, ACT 2601, Australia}
\affil{Department of Astronomy, The Ohio State University, Columbus, OH 43210, USA}
\affil{Center for Cosmology and AstroParticle Physics, The Ohio State University, Columbus, OH 43210, USA}
\author{Thor Tepper-Garc\'ia}
\affil{Sydney Institute for Astronomy, School of Physics, University of Sydney, NSW 2006, Australia}
\affil{Centre of Excellence for All-Sky Astrophysics in Three Dimensions (ASTRO-3D), Australia}
\author{Joss Bland-Hawthorn}
\affil{Sydney Institute for Astronomy, School of Physics, University of Sydney, NSW 2006, Australia}
\affil{Centre of Excellence for All-Sky Astrophysics in Three Dimensions (ASTRO-3D), Australia}
\author{Jeffrey Carlin}
\affil{Rubin Observatory/Legacy Survey of Space and Time (LSST), 950 N. Cherry Ave. Tucson, AZ, 85719, USA}

\correspondingauthor{HFW}
\email{haifeng.wang.astro@gmail.com};\\

\begin{abstract}

We present a novel, deep-learning based method -- dubbed Galactic-Seismology Substructures and Streams Hunter, or GS$^{3}$ Hunter for short, to search for substructures and streams in stellar kinematics data. GS$^{3}$ Hunter relies on a combined application of Siamese Neural Networks to transform the phase space information and the K-means algorithm for the clustering. As a validation test, we apply GS$^{3}$ Hunter to a subset of the Feedback in Realistic Environments (FIRE) cosmological simulations. The stellar streams and substructures thus identified are in good agreement with corresponding results reported earlier by the FIRE team. In the same vein, we apply our method to a subset of local halo stars from the Gaia Early Data Release 3 and GALAH DR3 datasets, and recover several, previously known dynamical groups, such as Thamnos 1+2, Hot Thick Disk, ED-1, L-RL3, Helmi 1+2, and Gaia-Sausage-Enceladus, Sequoia, VRM, Cronus, Nereus. Finally, we apply our method without fine-tuning to a subset of K-giant stars located in the inner halo region, obtained from the LAMOST Data Release 5 (DR5) dataset. We recover three, previously known structures (Sagittarius, Hercules-Aquila Cloud, and the Virgo Overdensity), but we also discover a number of new substructures. We anticipate that GS$^{3}$ Hunter will become a useful tool for the community dedicated to the search of stellar streams and structures in the Milky Way (MW) and the Local group, thus helping advance our understanding of the stellar inner and outer halos, and of the assembly and tidal stripping history in and around the MW.

\end{abstract}


\section{Introduction} 

The assembly history of the Milky Way (MW), which has formed over billions of years as the result of many accretion and merger events, is imprinted in its stellar substructures and streams \citep[e.g.,][]{2000AJ....119.2843C,2002ARA&A..40..487F,2004ASPC..317..256M,Heidi2016,2016MNRAS.463.1759B,2018MNRAS.478..611B,2018Natur.563...85H,2019A&A...631L...9K,2019ApJ...872..152I,2020ApJ...891...39Y,2020ApJ...901...48N,2022ApJ...926..107M}. 

In recent years, many research groups have discovered new Galactic stellar streams across different chemical and dynamical spaces as well as over a range of stellar orbital parameters \citep[e.g.,][]{2019MNRAS.490.3508L, 2022ApJ...928...30L, 2021hst..prop16791B, 2021ApJ...909L..26B}. \citet{2016ASSL..420...87G} have pinpointed the location and estimated the kinematic properties as well as the chemical abundances for some of the major Galactic streams such as the Sagittarius Stream, the Virgo Stellar Stream, and the Orphan Stream or Orphan-Chenab large stream (OC stream). Recently, \citet{2023MNRAS.520.5225M} published the {\em Galstreams} library, a uniform compilation of the orbital parameters (position and velocity) for nearly hundred known Galactic stellar streams. In addition, that study presents an analysis of the issues and uncertainties in the parameters of individual streams across the library, providing clear pointers of needed improvement. The library provides a convenient platform to catalogue known stellar streams, and promotes the addition and follow-up of new discoveries, greatly assisting the study and our understanding of the stellar streams and structures in and around the MW.

Theoretical models suggest that, under the hierarchical cosmological assembly process, accreting objects will retain their original orbital properties and kinematic states after structurally phase mixing \citep[e.g.][]{2000MNRAS.319..657H}. A long-standing problem is to determine which substructures (or streams) were originally formed in-situ (i.e. in the Galactic stellar halo), and which ones formed in external systems that were then  accreted onto the MW as part of later events. One possible solution to this is to estimate as accurately as possible the kinematic and orbital parameters of stellar substructures and streams around galaxies, since their orbital properties are effective tracers of the galaxy formation history and gravitational potential \citep{2010ApJ...714..229L}. Thus, the availability of such tracers is essential for our understanding of the background and history of galaxy formation, as well as the origin and birth environment of substructures and streams.

Streams can be formed when globular clusters or dwarf galaxies are tidally stripped (or entirely destroyed) as a result of the gravitational interaction with the MW. Some  disk substructures or disk streams can also form by internal and external perturbations \citep{wang2018b,wang2019,wang2020a,wang2020b,wang2020c,wang2023a}. A prime example of the former process are the Sagittarius stellar \citep[e.g.]{2010ApJ...714..229L, 2011Natur.477..301P}, and gaseous \citep{TepperGarcia2018} streams, which formed during the infall of the Sagittarius dwarf \citep{1994Natur.370..194I} onto the MW. One stream formed during the dwarf's first orbital wrap and is located at roughly $\tilde{\Lambda}_{\odot}$ $\in$ [$-$${180}^{\circ}$, ${180}^{\circ}$], the second at $\tilde{\Lambda}_{\odot}$ $\in$ [$-$${540}^{\circ}$, $-$${180}^{\circ}$] $\cup$ [${180}^{\circ}$, ${540}^{\circ}$] and the third at $\tilde{\Lambda}_{\odot}$ $\ge$ ${540}^{\circ}$ \citep{2022A&A...666A..64R}. Many models for stellar streams and their structure have also been continuously proposed and some colleagues have suggested that the streams provide an effective way to study the distribution of dark matter and potential in galaxies \citep[e.g.]{2022ApJ...940L...3W,2010MNRAS.408L..26P,2021MNRAS.501.2279V,2010ApJ...714..229L}.

A promising approach to identifying signatures of (Galactic) accretion events is based on the use of machine-learning (clustering) methods coupled to (ideally) comprehensive datasets. In this respect, large photometric and spectroscopic surveys such as the Sloan Digital Sky Survey \citep[SDSS;][]{2000AJ....120.1579Y}, the Two Micron All-Sky Survey \citep[2MASS;][]{2000AJ....119.2498J}, or the Large Sky Area Multi-Objective Fiber Optic Spectroscopic Telescope \citep[LAMOST;][]{2012RAA....12..723Z, Deng2012}, in addition to the full 6D phase-space information of millions of stars in the Solar Neighbourhood provided by the Gaia satellite, in particular its (E) Data Release 3 \citep[GDR3 or GEDR3;][]{2022arXiv220605902K}, provide an exquisite dataset to which these clustering techniques can be seamlessly applied.

The central idea of any these methods is to search for significant clusters in the distribution of stars in a suitable projection space. For instance, some earlier studies have looked at the distribution of stars in six-dimensional (6D) phase space, or on a plane defined by the corresponding integrals of motions such as total energy ({\tt\emph E}), the component of angular momentum parallel to the {\tt\emph z} axis ({\tt\emph L$_z$}), and the in-plane component of angular momentum {\tt\emph L$_\bot$} $( = \sqrt{{\tt\emph L_x}^2 + {\tt\emph L_y}^2})$ \citep{2022A&A...665A..57L}, while other studies have made use of action-angle coordinates to detect the clustering signal of streams and substructures based on ENLINK algorithm \citep{2009ApJ...703.1061S, 2021AAS...23755209W}.

Previous detection methods for stellar streams include the matched filter (MF) technique \citep{2002AJ....124..349R,2011MNRAS.416..393B}; the detection of co-moving groups of stars \citep{2013ApJ...765L..39M}; and the Pole Count technique \citep{1996ApJ...465..278J}. Some comparisons for these three methods may be found in \citet{2018MNRAS.477.4063M}.

Other clustering algorithms, developed and improved over many years, have been used to find and identify streams and substructures \citep{2018MNRAS.477.4063M,2018ApJ...863...26Y,1985ApJ...292..371D,2022A&A...665A..57L}. These include:

(i) FOF: The FOF (friends-of-friends) algorithm \citep{1985ApJ...292..371D}, is a classical algorithm for detecting (sub)structures. It is also a purely geometric method in which particles that are closer than a given scale (link length) calculated by Euclidean distance are connected. Different particles will be distributed in different regions. This method needs identifying substructures within the halo at a fixed linkage setting. Using the LAMOST DR5 K-giants sample with distances {\tt\emph d} $\approx 5 -120$ kpc and FOF methods, \citet{2019ApJ...880...65Y} have identified streams and substructures in the Galactic halo, such as the Sagittarius Streams, the Monoceros Ring, the Virgo Overdensity, the Hercules-Aquila Cloud, the Orphan Streams, and other, previously unknown substructures.

(ii) StarGO (Star's Galactic Origin): Developed by \citet{2018ApJ...863...26Y}, it is a substructure identification method based on the self-organizing map (SOM) technique \citep{2001som..book.....K}. The advantage of the algorithm is that the SOM grid automatically finds the nearest node and updates the position based on iteration and Euclidean distance calculation, and has successfully identified structures such as the Cetus stream \citep{2009ApJ...700L..61N}, amongst others. SOM is used to generate a low-dimensional space for training samples, which can transform complex nonlinear statistical relationships among high-dimensional data into simple geometric relationships and display them in a low-dimensional way to provide a convenient way for further calculations. And the aim of SOM is to map an n-D input data to a 2D neural map while retaining the topological structures within the data at the same time. The main point we should note about SOM is that this algorithm needs to provide weight information to a large quantity of neurons in order to cluster the input data. To be successful, the input to the weight layer must be  carefully chosen. Using StarGO, the metal-poor outer part of stellar halo of the Milky Way ({\tt\emph d} $>$ 15 kpc) and the northern counterpart of the Cetus structure were unravelled in \citet{2019ApJ...881..164Y}.

(iii) Single linkage cluster: The single linkage algorithm \citep{2022A&A...665A..57L, 2022A&A...665A..58R} is a hierarchical clustering method for detecting (sub)-structures. The distance between the nearest two nodes is calculated using the Mahalanobis distance. Applying this technnique, \citet{2022A&A...665A..57L} found 67 local halo clusters in the integral-of-motion space within {\tt\emph d} = 2.5~kpc from the Sun. This algorithm is computationally intensive because it has to calculate the two-by-two distance of all data points in multiple clusters each time. In addition, since hierarchical clustering uses the greedy algorithms, one of the characteristics of such algorithms is that the results obtained may be locally optimal and not necessarily globally optimal. But this is not necessarily a disadvantage, because the locally optimal solution is also sufficient to solve the problem in the search for agglomerative structures.

\citet{2023A&A...670L...2D} have adopted the Single-linkage clustering method proposed by \citet{2022A&A...665A..57L} and found 7 main dynamic substructures in the local Galactic stellar halo, some of which are believed to have formed in-situ, and others which likely formed during earlier accretion events. Most of the identified groups belong to larger structures such as the Gaia-Enceladus-Sausage \citep[GSE;][]{2018Natur.563...85H, 2018MNRAS.478..611B, 2003ApJ...585L.125B}, the L-RL3 (\citet{2023A&A...670L...2D}\footnote{We adopt the numenclature introduced by \citet{2023A&A...670L...2D}.}, the Hot Thick Disk \citep{2019A&A...632A...4D, 2018Natur.563...85H}, Sequoia \citep{2018ApJ...860L..11K}, the Helmi Stream \citep{2022arXiv220513810R}, and Thamnos \citep{2018ApJ...860L..11K}. In addition, \citet{2023A&A...670L...2D} discovered a previously unknown substructure -- dubbed ED-1, and identified 11 independent clumps, five of which are previously unknown (ED-2, 3, 4, 5 and 6).

\citet{2012MNRAS.426L...1A} have applied the clustering method of wavelet transform to identify the overdensity distribution in velocity space around the Sun. They found that the main kinematic groups are consistent with belonging to large-scale features, and they also identified a new group in the local disk region.

With the help of the hierarchical clustering method and the Gaia early DR3 (EDR3) dataset, as well as catalogs from ground-based spectroscopic surveys, using a single linkage algorithm defined by the usual integration of the motion energy {\tt\emph E} and the two components of the angular momentum {\tt\emph L$_z$} and {\tt\emph L$_\bot$}, \citet{2022A&A...665A..57L} actually unveiled 6 main groups or substructures, more population properties of these substructures were also discussed in a later work \citep{2022A&A...665A..58R}. In contrast, \citet{2023ApJ...944..169D} have also analyzed the components in the local halo using the Bayesian Gaussian mixed model regression algorithm to characterize the local stellar halo, and they found that the data fit the model best with four components, which are Virgo Radial Merger (VRM), Cronus, Nereus, and Thamnos. They suggest that the GSE merger event is probably a combination of these four components.

The Stream Finder method \citep{2018MNRAS.477.4063M}, has been developed to explicitly identify dynamically cold and thin stream structures based on their orbit parameters. Using this method, \citet{2018MNRAS.477.4063M} have succeeded in detecting the GD-1 stellar stream based on the Pan-STARRS1 dataset \citep{2018MNRAS.478.3862M}. Applying this method to the Gaia DR2 dataset, \citet{2020ApJ...891L..19I} have presented a complete six-dimensional panorama of the Sagittarius stream.
Moreover, \citet{2022ApJ...930..103Y} combined the StarGO method together with Stream Finder and found two structures in the southern extension of the northern Cetus stream: the Palca stream, and a new southern stream that wraps around the sky extending over $100^{\circ}$. Moreover, by combining datasets of 70 globular clusters, 41 streams, and 46 satellites together, \citet{2022ApJ...926..107M} detected a total of 6 groups, including the known Sagittarius, Cetus, Gaia-Sausage/Enceladus, LMS-1/Wukong, Arjuna/Sequoia/I$^{'}$itoi and a new merger event they refer to as Pontus, also with the help of ENLINK software \citep{2009ApJ...703.1061S}.

In this paper, we introduce a new concept for the identification of stellar streams or substructures based on a novel clustering technique. Our method relies mainly on the application of the Siamese Neural Networks \citep{2018PatRe..74...52S} to transform the 6D phase-space information upon which the groups of structures (substructures) are identified by a K-means clustering algorithm. %Since this method calculates the distance in the kinematic parameter space, a potential drawback of the method is a reduced sensitivity to structures that are more diffusely distributed in space, but as we demonstrate below, this is not an issue.

This paper is organized as follows: In Section \ref{section:data}, we describe the dataset selection and processing required before it is subject to our analysis. In Section \ref{section:method} we describe the technical details of our novel algorithm. In Section \ref{section:result} we present some properties of the stellar structures we find by applying our algorithm to the dataset, mainly including velocity space. We also present the structures identified in the inner halo as a by-product of applying our method to the LAMOST dataset, and show their properties and spatial distributions. In Section \ref{section:discussion}, we summarize our results and compare them with the results from other groups obtained by means of other methods. In Section \ref{section:Conclusion} we conclude our discussion.

\section{Data} 
\label{section:data}

\begin{figure*}[!ht]
  \centering
  \includegraphics[width=7.5in]{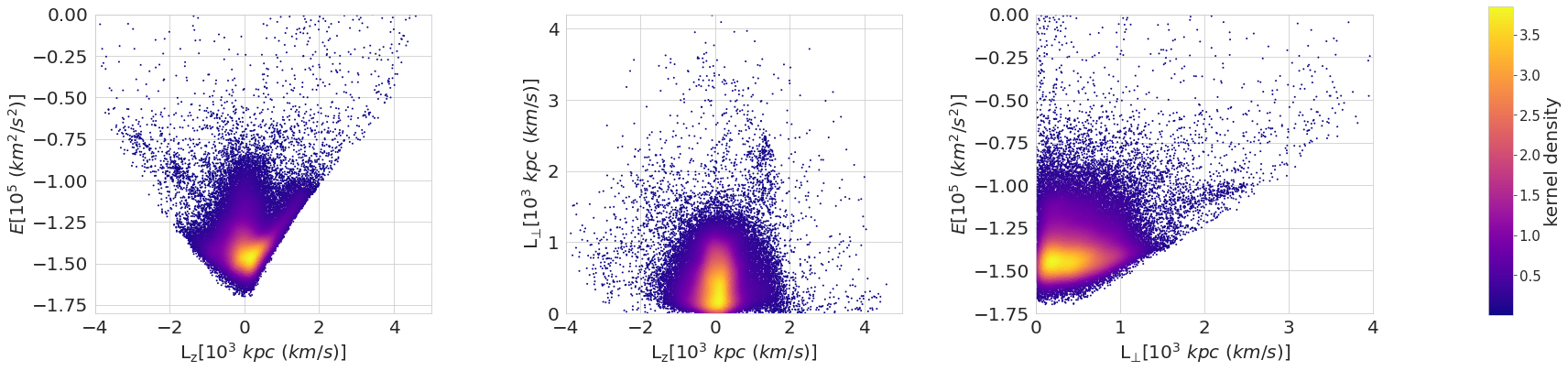}
  \caption{Local halo sample distribution in integral of motion space (angular momentum and energy), coloured by the normalized star counts. The sample includes 51,671 stars in the solar neighborhood with 6D phase space information.}
  \label{sun(EL)}
\end{figure*}

% figure 1 + For a detailed explanation of the data selection, thanks to the work of \citet{2022A&A...665A..57L}, here we just have a reconstruction and then use our novel algorithm to find groups.

\begin{figure*}[!ht]
  \centering
  \includegraphics[width=7.5in]{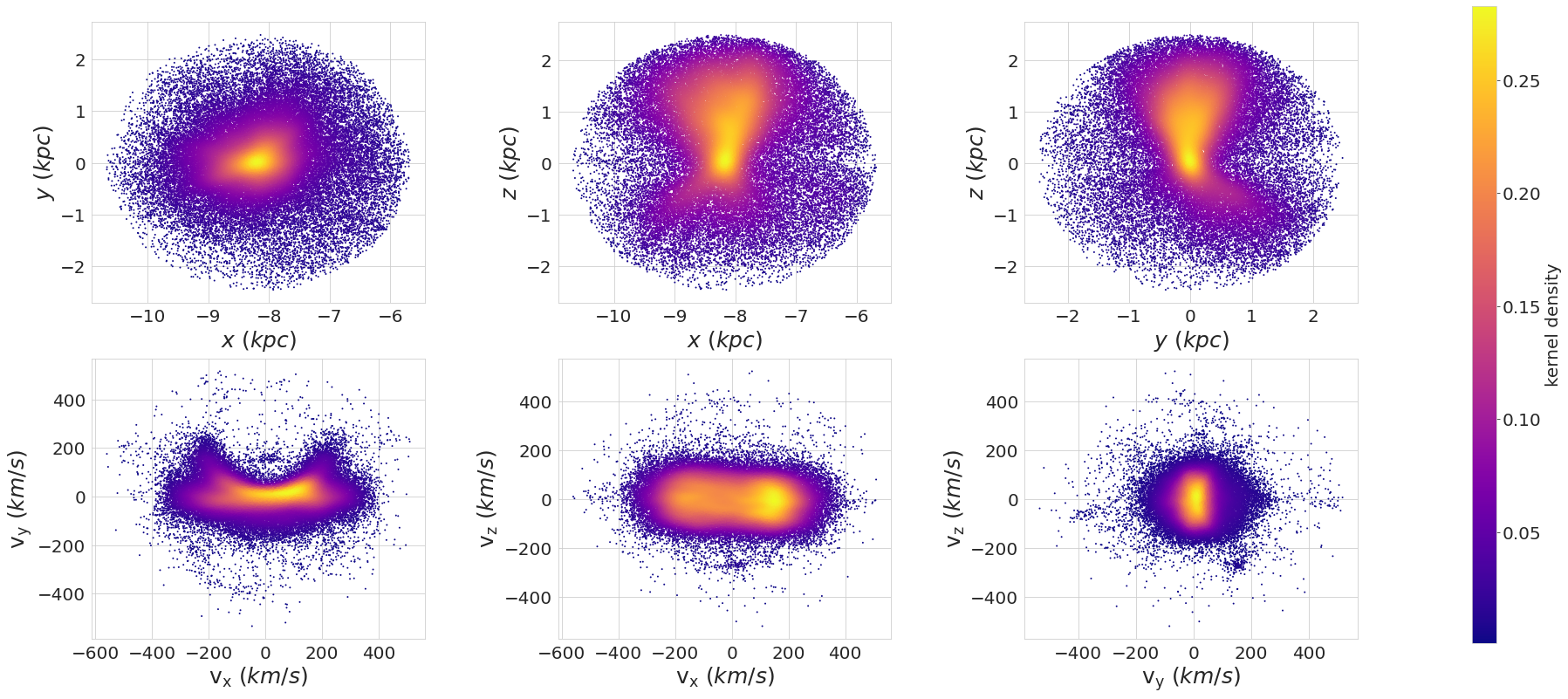}
  \caption{Local halo sample distribution in the 6D phase space as a complement of Figure \ref{sun(EL)} in Galactic coordinates.}
  \label{localhalopositiongalactic}
\end{figure*}

\subsection{Dataset for the local halo structures (also belonging to Disk Region)} 

\begin{figure}[!t]
  \centering
  \includegraphics[width=3.5in]{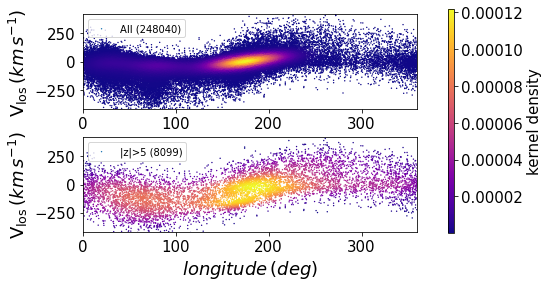}
  \caption{Galactic longitude versus line-of-sight velocity of LAMOST K giants. The top panel shows the whole sample of 248,040 stars crossed match with Gaia DR3 within 1$^{\prime \prime}$. The bottom panel is the sample $\mid {\tt\emph z} \mid$ $ > $ 5 kpc, the edge-on projection is shown in Figure \ref{X_Z_plot}.}
  \label{glon_rv_plot}
\end{figure}

For the local halo search, we utilized the dataset compiled by \citet{2022A&A...665A..57L}, which incorporates Gaia EDR3 RVS \citep{2021A&A...649A...1G} sample of stars with their line-of-sight velocities measured by the Radial Velocity Spectrometer, supplemented by radial velocities observed by ground-based spectroscopy such as the sixth data release of the Large Area Multi-Objective Fiber Optic Spectroscopic Telescope (LAMOST DR6) low-resolution \citep{2020ApJS..251...27W} and medium-resolution survey \citep{2019RAA....19...75L}, the sixth data release from the RAdial Velocity Experiment \citep[RAVE DR6;][]{2020AJ....160...82S}, the third data release from Galactic Archaeology and HERMES \citep[GALAH DR3;][]{2021MNRAS.506..150B}, and the 16th data release from the Apache Point Observatory Galactic Evolution Experiment \citep[APOGEE DR16;][]{2020ApJS..249....3A}.

Stars with distances greater than 2.5 kpc have been removed from this catalog; the distance was computed by inverting the parallaxes after correcting for a zero-point offset of 0.017 mas {\bf\citep{2021A&A...649A...2L}}. To improve the quality of the data, stars for which the normalized unit weight error is larger than 1.4 are removed. For this work, the local standard of rest {\tt\emph V$_{LSR}$} = 232 km s$^{-1}$, the distance of the Sun from the Galactic center is 8.178 kpc \citep{2019A&A...625L..10G}, and ({\tt U$_{\odot}$}, {\tt V$_{\odot}$}, {\tt W$_{\odot}$}) = (11.1, 12.24, 7.25) km s$^{-1}$ is the peculiar motion of the Sun \citep{2020ApJ...892...39R, 2010MNRAS.403.1829S}, all of which are consistent with the Galaxy parameter set recommendations of \citet{2016ARA&A..54..529B}.

\begin{figure}[!t]
  \centering
  \includegraphics[width=0.4\textwidth]{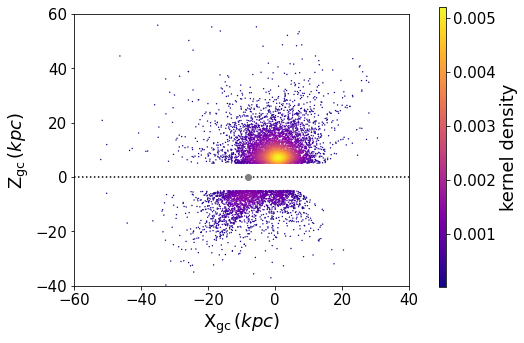}
  \caption{Spatial distribution on the {\tt\emph x}$-${\tt\emph z} plane for 8099 LAMOST inner halo K giant stars, which will be used to for a quick investigation of the inner halo substructures, as a by-product of this work. The K giant catalog and selection criteria are based on the \citet{2014ApJ...790..110L} and \citet{2019ApJ...880...65Y}, respectively. More detailed inner and in particular outer halo streams will be shown in future work.}
  \label{X_Z_plot}
\end{figure}

Here the total energy is: {\tt\emph E} = {\tt V$^{2}/2$} + ${\tt \Phi(r)}$, where {\tt\emph V} is the the total velocity, ${\tt \Phi(r)}$ is the gravitational potential of the galaxy at the location of the star, and the potential is calculated from the \citet{2019A&A...625A...5K} and \citet{2022A&A...665A..57L}, that is, the potential includes a Miyamoto-Nagai disk, Hernquist bulge, and NFW halo. The Milky Way parameters are ({\tt\emph a$_d$}, {\tt\emph b$_d$}) = (6.5, 0.26) kpc, {\tt\emph M$_d$} = 9.3 $\times$ $10^{10}$ M$_{\odot}$ for the disk, {\tt\emph c$_b$} = 0.7 kpc, {\tt\emph M$_b$} = 3.0 $\times$ $10^{10}$ M$_{\odot}$ for the bulge; and {\tt\emph r$_s$} = 21.5 kpc, {\tt\emph c$_h$} = 12, {\tt\emph M$_{halo}$} = $10^{12}$ M$_{\odot}$ for the halo (after tests we find that different mass will not change our final conclusions). More details can be found in \citet{2022A&A...665A..57L}. As mentioned in \citet{2022A&A...665A..57L}, the halo stars are selected by requiring $|${\tt\emph V} - {\tt\emph V$_{LSR}$}$|$ ${\textgreater}$ 210 km s$^{-1}$.

\begin{figure*}[!ht]
  \centering
  \includegraphics[width=6in]{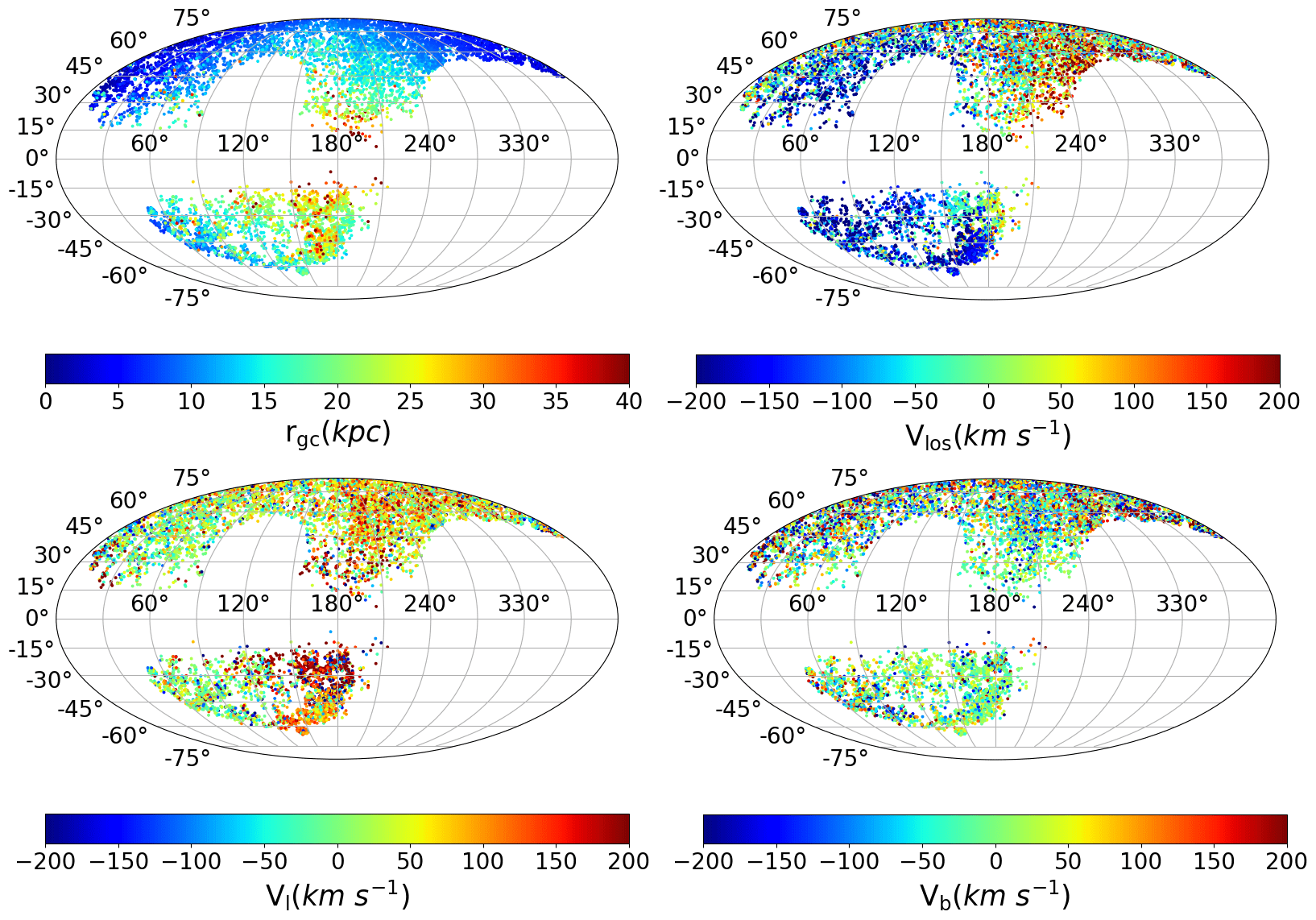}
  \caption{Galactic sky coverage of the LAMOST final halo K-giants sample we adopt in this work. Stars are colored according to Galactocentric distance {\tt\emph r$_{gc}$}, line-of-sight velocity {\tt\emph V$_{los}$}, and tangential velocities ({\tt\emph v$_l$}, {\tt\emph v$_b$}).}
  \label{lb_plot}
\end{figure*}

We make full use of this publicly available dataset with {\tt\emph N} = 51,671 stars in our Deep Learning methods\footnote{We adopt EDR3 for this project as a proof of concept because this work was initiated during the time when only EDR3 is available.}. Figure \ref{sun(EL)} shows the recovered sample distribution for the local halo stars in IOM space, colored according to the density. As seen, the energy peak is around $-$1.5 $\times 10^5$ (km$^2$/s$^2$) and the vertical angular momentum peak is around 0 $\times10^3 $ kpc km/s. The spatial and Galactic velocity distribution are shown in the Figure \ref{localhalopositiongalactic}. %Our tests using a range of Galactic masses show that different potentials do not significantly affect our final results.

\subsection{Dataset for inner halo region}

%We apply our method to Gaia DR3 radial velocity sample (RVS) and LAMOST DR5 K giants to identify substructures/structures/streams for the inner halo, and also as a way to test the performance of our method.

%Gaia DR3 \citep{2022arXiv220605989B} measured a total of 1.8 billion sources, and enabled a significant number of new datasets, many of which are updates to similar datasets produced from Gaia EDR3.  For example, Gaia DR3 includes updates to the mean radial velocities, chemical information, etc.

For the inner halo streams search, the dataset is based on the LAMOST DR5 catalog of about 9 million spectra, of which about 5 million have information on stellar parameters and radial velocities. Using the criteria mentioned in \citet{2014ApJ...790..110L,wang2018b}, we select a sample of ~1 million K giant stars. After cross-matching with Gaia DR3 using TOPCAT \citep{2005ASPC..347...29T} with a matching radius of 1$^{\prime \prime}$, we obtain detailed information about the sky position, distance, line-of-sight velocity and proper motion of 248,040 K giant sources after cross matching with Gaia DR3 dataset. Figure \ref{glon_rv_plot} (upper panel) shows the distribution of the line-of-sight velocity with longitude for the 248,040 K giants we have selected, most of which are located in the disk.

Since we wish to focus on the Galactic inner halo, we exclude K giants within 5 kpc of the Galactic disk ($|${\tt\emph z}$|$ ${\textless}$ 5 kpc) according to the selection criteria in \citet{2019ApJ...880...65Y}. %We assume that the Sun is at distance $R_{\odot}$ = 8.178 kpc from the Galactic center \citep{2019A&A...625L..10G} and $z_{\odot}$ = 0.02 kpc above the Galactic plane \citep{2019MNRAS.482.1417B}. We use a solar motion of ($v_{x_{\odot}}$, $v_{y_{\odot}}$, $v_{z_{\odot}}$) = (-11.1, -248.5, 7.25) km s$^{-1}$ \citep{2020ApJ...892...39R,2010MNRAS.403.1829S}. 

After removing stars with $|${\tt\emph z}$|$ ${\textless}$ 5 kpc, most of the disk stars are excluded, as shown in the lower panel of Figure \ref{glon_rv_plot}. Finally, we build a sample of 8099 halo K giants with 3D position, 3D velocity and metallicity. The spatial distribution of our final sample in the {\tt\emph x}-{\tt\emph z} plane is shown in Figure \ref{X_Z_plot}, and the sky distribution of halo K-giants is also displayed in Figure \ref{lb_plot}. Most of the K-giants have Galactocentric distances between 5$-$60 kpc. Next, we will detect and identify substructures from the selected sample of halo K-giants using the GS$^{3}$ Hunter method, which will be demonstrated in the results section.

\section{Methodology and Validations}
\label{section:method}

In our study, we tackle the challenge of analyzing data within a complex 6-dimensional (6D) kinematic space. The inherent sparsity and irregularity of this data demand a sophisticated approach for effective clustering. Our methodology integrates neural network architectures and clustering techniques to simplify the data and enhance its interpretability.

At the heart of our method is the Siamese Network \citep{2018PatRe..74...52S}, an architecture adept at comparing and identifying relationships between pairs of input data. This network is ideal for our purpose as it processes two parallel input streams, learning to discern the similarities or dissimilarities between them. Specifically, the Siamese network is trained to map the input data into some embedding space {\tt\emph z}, and dissimilar pairs are farther apart. Next, by calculating the distance of data pairs in the embedding space, their similarity can be determined. Two data points, {\tt\emph x$_0$} and {\tt\emph x$_1$}, are considered positively (similar) related if their Euclidean distance is below a specific threshold ({\tt siam$_D$}) in the embedding space. Our selection process for the siam$_D$ value involved: $a).$ Referencing the Mahalanobis distance results from the previous step. $b).$ Conducting multiple tests to verify that the final number of candidate groups met our filtering criteria. $c).$ Empirically choosing the most appropriate siam$_D$ value based on these tests. While we aim to make this parameter more flexible in future iterations of our method, we believe that a fixed value is reasonable for the current study. The sensitivity of our results to this parameter choice is an important consideration, and we acknowledge that a more detailed analysis of this relationship could provide valuable insights into the robustness of our method.

% Specifically, our Siamese Network encodes an input $\mathbb{R}^6$ (representing the 6D phase space) into a 2D output \( z \in \mathbb{R}^2 \). Each data point \( x \in \mathbb{R}^6 \) is transformed into \( z \in \mathbb{R}^2 \), aiming to preserve the relational dynamics in a reduced dimensionality space.

To quantify the similarity of the data points in the input parameter space and get the value of siam$_D$, we used the Mahalanobis distance\footnote{The relevant documentation can be found at \url{https://docs.scipy.org/doc/scipy/reference/generated/scipy.spatial.distance.mahalanobis.html}}, compared with the Euclidean distance, the Mahalanobis distance takes more into account the measurement covariances between different features. The Mahalanobis distance is calculated as:

\begin{equation}\label{2}
\tt{D_M(x_0,x_1)= \sqrt{(x_0-x_1)^{T} \Sigma^{-1} (x_0-x_1)}},
\end{equation}

\noindent
In our case, the 6D covariance ${\tt \Sigma}$ is derived from the Gaia phase space data.

Training the Siamese Network focuses on minimizing a contrastive loss objective that draws similar data points closer in the embedding space, while distancing the dissimilar ones. The loss function (which we refer to \citet{2018arXiv180101587S}) is:

\begin{equation}
\tt{L(x_0, x_1)}=
\left\{
    \begin{array}{lc}
        ||z_0 - z_1||^2, & (x_0, x_1)\ is \ positive \\
        max(c - ||z_0 - z_1||, 0)^2, & (x_0, x_1)\ is \ negative \\
    \end{array}
\right.
\end{equation}

\noindent
where c is a margin (typically set to 1), Siamese net maps and the data point {\tt\emph x$_0$} into the embedding space {\tt\emph z$_0$} and clusters closely related samples in {\tt\emph z} space within a specified threshold ({\tt siam$_D$}). The distance in the {\tt\emph z} space is assumed to be Euclidean. The input {\tt\emph x} is in $\mathbb{R}^6$ which consists of the phase space information. {\tt siam$_D$} here serves to generate pseudo labels for the constrastive learning since the data is unlabelled.

The architecture of the Siamese network follows closely from \citet{2018arXiv180101587S}. Briefly, in this work it consists of an MLP of 4 layers, with the activation of RELU (Rectified Linear Unit) function, which is a commonly used activation function in neural networks. It introduces non-linearity into the network, enhancing its capacity to learn complex patterns from data. And a Batch Normalization layer has been added to normalize the data and accelerate network convergence. Additionally, a custom Lambda layer has been included to compute the Euclidean distance in the embedding space {\tt\emph z}.

To further enhance the constrast in the clustering especially in the high density region, we further feed the output of the Siamese network into a Multi-Layer Perceptron (MLP), mapping into the final transformed space of $y \in \mathbb{R}^s$. The MLP has a structure consisting of four layers with a relu activation function, followed by a layer with a tanh activation function. We also assume the L2 regularization which improves generalisation and reduces overfitting during training. And the MLP trained with a self-supervision loss:

\begin{equation}
L(\theta) = \mathbb{E}_{(x_i,x_j)} \left[ \omega(z_0,z_1) ||y_0 - y_1||^{2} \right]
\end{equation}

\noindent
where $\omega(z_0, z_1)$ is the affinity matrix calculated by KNN\footnote{The relevant documentation can be found at \url{https://scikit-learn.org/stable/modules/generated/sklearn.neighbors.KNeighborsClassifier.html}} in the embedding space, $||y_0 - y_1||$ is the similarity between $y_0$ and $y_1$ calculated by the Euclidean distance. ($y_0$, $y_1 \in \mathbb{R}^s$). The expectation is done over all pairs, although due to the affinity matrix, only few pairs contributed. The affinity matrix of the embedding is used as the weights for the loss function in order to put more weights on ``nodes" with the higher density, and therefore require the network to pay more focus on these challenging regime. In practice, we found that by setting $s$ to be the number of clusters is a good rule of thumbs.

Finally, the output embedding ${\tt (y \in \mathbb{R}^s)}$ undergoes unsupervised clustering through the K-means algorithm\footnote{The relevant documentation can be found at \url{https://scikit-learn.org/stable/modules/generated/sklearn.cluster.KMeans.html}}, an empirically determined number of clusters sufficient to capture structures in the $\mathbb{R}^s$ embedding space. 

% At the end, we receive a result labeled one-to-one with the input data, where samples with the same label are considered as a cluster/group.

We note that, K-means clustering can produce minor, non-robust clusters. To mitigate this, we introduce a density-based validation step using the Density Peaks Clustering Algorithm \citep[DPCA;][]{2014Sci...344.1492R}. For each cluster identified by K-means, we determine the peak via DPCA. From this peak, we define a radius $r = \sigma \times \sqrt{\lambda}$, where \( \sigma \) represents a confidence interval based on the $\chi^2$ distribution (\( \chi^2 = 12.95 \)). The choice of the $\chi^2$ distribution is motivated by the consistence of the algorithm compared to a $\chi^2$ distribution with six degrees of freedom. And \( \lambda \) is the eigenvalue from PCA performed on the cluster, considering the cluster's orientation in the 6D space. A cluster is deemed robust and significant if it contains 95\% of its data points within this radius.

This methodology presents a rigorous yet accessible strategy for interpreting data from a high-dimensional space. It offers a distinct advantage over classical methods like Friends-of-Friends (FoF), which primarily rely on Euclidean distance in feature space. Our use of the Siamese Network enables effective initial data transformation. We employ a machine-learning-driven method to automatically determine appropriate similarity through Siamese network training and map the originally loose and irregular structure in the initial data space into more isotropic groups in the $\mathbb{R}^s$ space by neural network transformation. Crucially, we also account for the inherent covariance among the 6D input parameters by initially assessing similarity between data points using the Mahalanobis distance. Additionally, GS$^{3}$ Hunter is proficient at detecting both dynamically cold and moderately hot stellar streams. When combined with sophisticated clustering and validation techniques, our method is well-equipped to uncover meaningful patterns and structures within the kinematic-dynamic space. This multifaceted approach not only ensures the robustness and relevance of the identified clusters over traditional clustering methodologies.

\begin{figure*}[!ht]%调节图片位置，h：浮动；t：顶部；b:底部；p：当前位置
  \centering
  \includegraphics[width=0.9\textwidth]{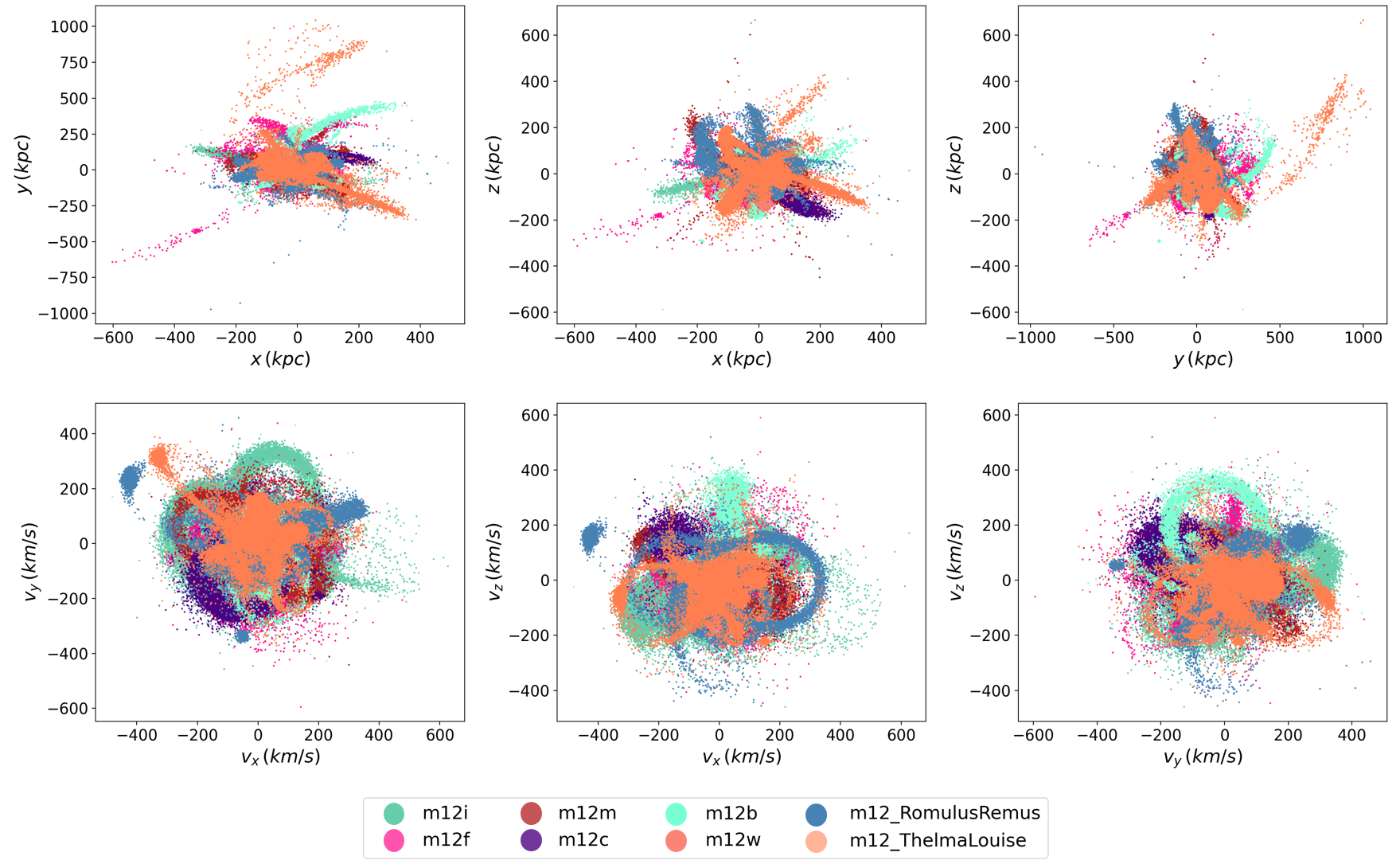}
  \caption{The spatial distribution of positions and velocities of the FIRE simulation data, where the different colors represent different labeled structures \citep{2021ApJ...920...10P}, more details could also be found in \citet{2016ApJ...827L..23W} and \citet{2023ApJS..265...44W}.}
  \label{FIRE_origin}
\end{figure*}

\begin{figure*}[!h]
  \centering
  \includegraphics[width=0.9\textwidth]{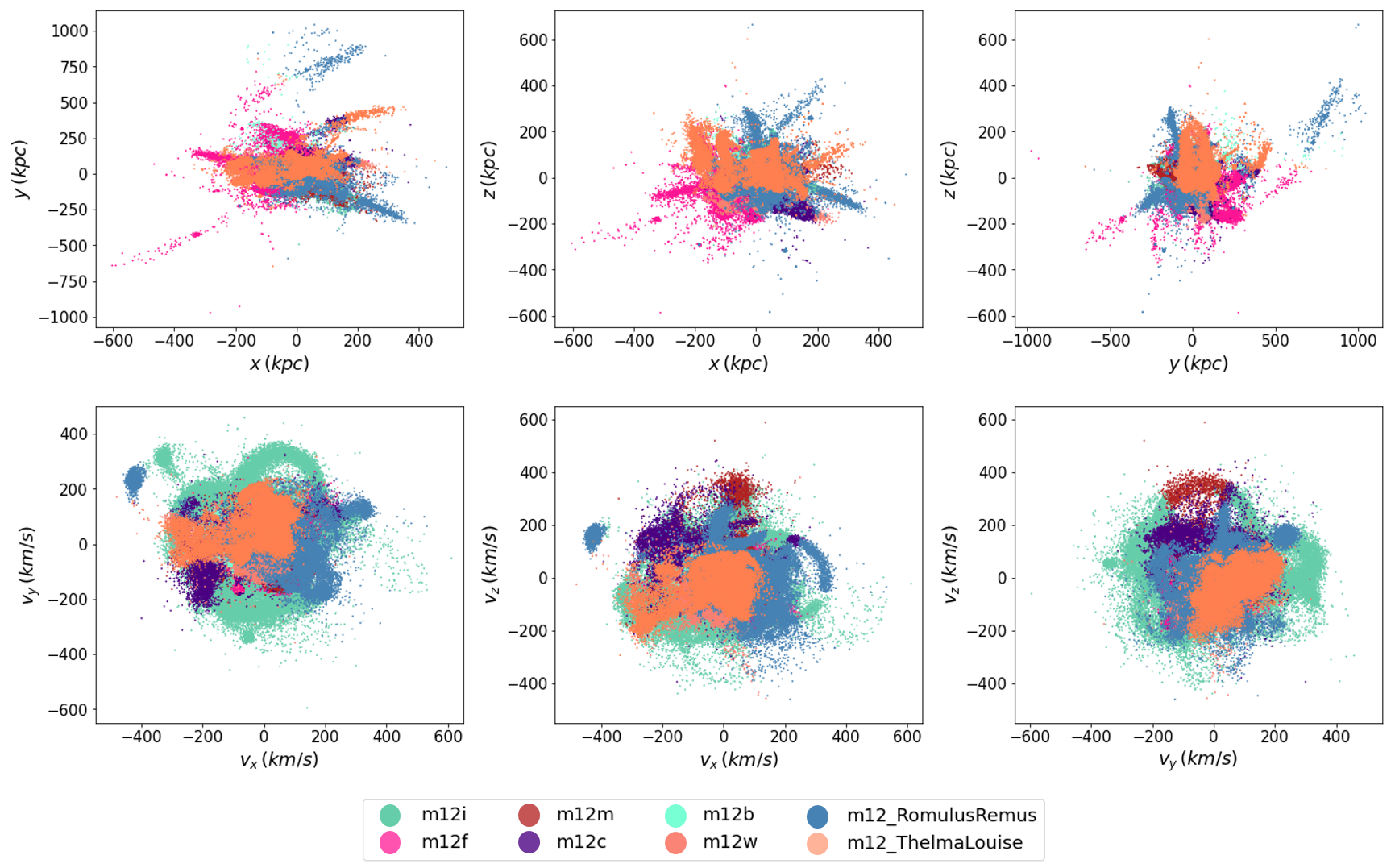}
  \caption{The results after applying GS$^{3}$ Hunter in FIRE are shown in 6D space, where the different colors correspond to the different structures shown the labels below.}
  \label{FIRE_GSHunter}
\end{figure*}

\begin{figure*}[!t]
  \centering
  \includegraphics[width=0.50\textwidth]{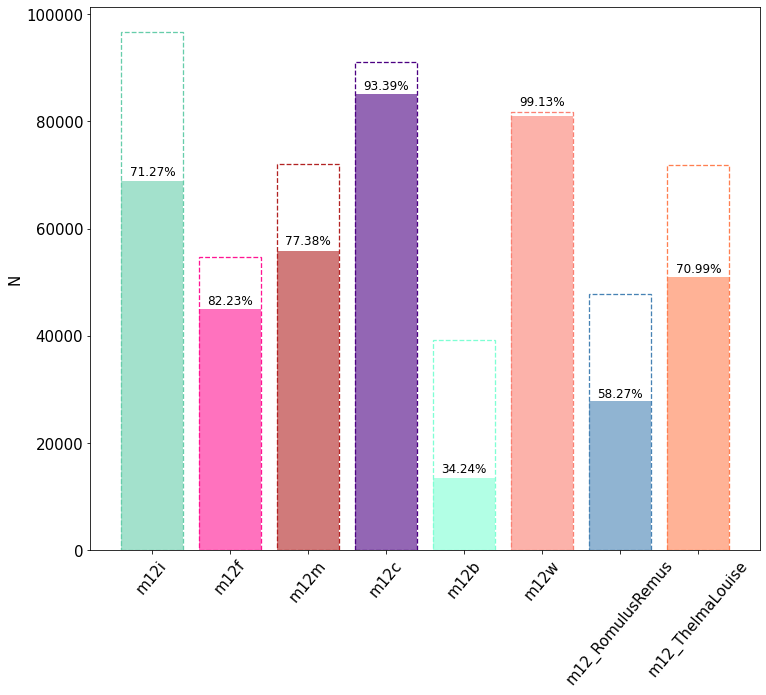}
  \caption{The fraction analysis for the results of Fig. \ref{FIRE_origin} and Fig. \ref{FIRE_GSHunter}. The horizontal coordinate is the name of the real structure in FIRE and the vertical coordinate is the number. The dashed part of the bar graph is the total number of each real structure, while the colored bars indicate the corresponding part of FIRE found by GS$^{3}$ Hunter. We labeled each bar with the percentage of the corresponding structure in the results. The low fraction for few mock galaxies (e.g., m12b) might be caused by the relatively higher stellar mass and velocity dispersion and phase mixing time in FIRE simulations, and the grouping classification of FIRE-2, in addition, the drawback of the method is another factor, which is not good at finding the relatively diffuse or very hot streams.}
  \label{FIRE_fraction}
\end{figure*}

We have tested the algorithm using the FIRE simulation\footnote{\url{https://fire.northwestern.edu/milky-way/}}, mainly from the FIRE-2 “Latte” and “ELVIS” suites. These simulations, detailed in previous literature \citep{2016ApJ...827L..23W, 2018MNRAS.480..800H, 2019MNRAS.487.1380G, 2020MNRAS.491.1471S, 2021ApJ...920...10P, 2022ApJ...939....2A, 2023ApJS..265...44W, 2023ApJ...949...44S, 2024ApJ...966..108N}, offer a rich dataset for our purpose. The “Latte” suite includes simulations of galaxies such as m12i, m12f, m12m, m12b, m12c, and m12w, while the “ELVIS on FIRE” suite encompasses the m12 Thelma \& Louise, and m12 Remus \& Romulus simulations. These simulations implement stellar evolution and feedback models, including processes such as stellar winds, core-collapse, Ia supernovae, and photoelectric heating, etc. The FIRE-2 simulations achieve parsec-scale resolution. In the work they have released three different suites as well as the accompanying (sub)halo catalogs of structures, and the FIRE data are currently publicly available\footnote{\url{http://flathub.flatironinstitute.org/fire}.} and more other analysis and details could be found in \citet{2023ApJS..265...44W}\footnote{\url{https://wetzel.ucdavis.edu/fire-simulations/}}. Our choice of the FIRE sample is influenced by its dominance in satellite and phase-mixed substructures, as classified by \citet{2021ApJ...920...10P}. Satelite is defined as a candidate with 120 - 10$^5$ star particles and maximum pairwise distances less than 120 kpc. Phase-mixed is defined as a candidate with 120 to 10$^5$ star particles, with the maximum pairwise distances exceeding 120 kpc, and a median local velocity dispersion above the value specified in their Eq.2. Stream is defined as a candidate with 120 to 10$^5$ star particles, with the maximum pairwise distances over 120 kpc, and a median local velocity dispersion below the value specified in their Eq.2. The differences in criteria for defining structures could explain some differences in classification. Specifically, we apply the GS$^{3}$ Hunter algorithm to the simulation data, which resulted in the identification of 33 distinct groups. Notably, among these groups, eight true progenitors were recognized. Regarding the case of multiple clusters being linked to a single structure, it's normal results of this work. The distribution of the original FIRE structures in Figure \ref{FIRE_origin}, Then we apply the GS$^{3}$ Hunter algorithm to the FIRE data, the effectiveness of the GS$^{3}$  Hunter algorithm in recovering these progenitors is detailed in Figure \ref{FIRE_GSHunter}, the fraction shown in Figure \ref{FIRE_fraction}. 

We also note there are some differences between the original mock data and the reconstructed results, which can be attributed to the following factors: our method utilizes 6-dimensional kinematic parameters (positions and velocities) as input. According to \citet{2021ApJ...920...10P}, phase-mixed structures may appear more dispersed in this space, affecting our results. Additionally, the choice of input space significantly impacts our results. The cutoff distance (Siam$_D$), which determines if two sample particles belong to the same group/cluster, is fixed. Consequently, simulations with more of phase-mixed structures and stellar streams will result in relatively poorer reconstructions using our method. 

Additionally, in FIRE simulations, higher-mass progenitor structures tend to become phase-mixed with the host galaxy more rapidly. Structures with larger masses generally exhibit greater velocity dispersion \citep[see Fig.10, 11 \& 12 in][]{2021ApJ...920...10P}, which can challenge our method's ability to detect diffuse structures. Considering these factors, the current results are reasonable for these differences.

There are two key observations from this exercise. First, the GS$^{3}$ Hunter algorithm successfully recovers the majority of the substructures, underscoring the robustness of our method. This achievement demonstrates the algorithm's capability in accurately identifying complex structures within the simulated data. Second, it's important to note that the eight identified progenitors were found across different groups. This outcome suggests that, while the downstream MLP network aids in grouping data closer to their true progenitors, the process is not flawless. The groups recovered by the GS$^{3}$ Hunter algorithm can, and in some cases do, originate from different progenitors. This observation highlights the nuanced nature of the data and the complexity involved in perfectly aligning the algorithmically identified groups with their true progenitors. 

We also show the results for the detailed fraction in Table \ref{FIRE} in the Appendix. Next, we will compute and analyze the datasets for groups finding from GS$^{3}$ Hunter, and we also summarize the resulting results in Table \ref{tab:2} and Table \ref{tab:3} presented separately in the Appendix.

\begin{figure*}[!t]%调节图片位置，h：浮动；t：顶部；b:底部；p：当前位置
  \centering
  \includegraphics[width=0.95\textwidth]{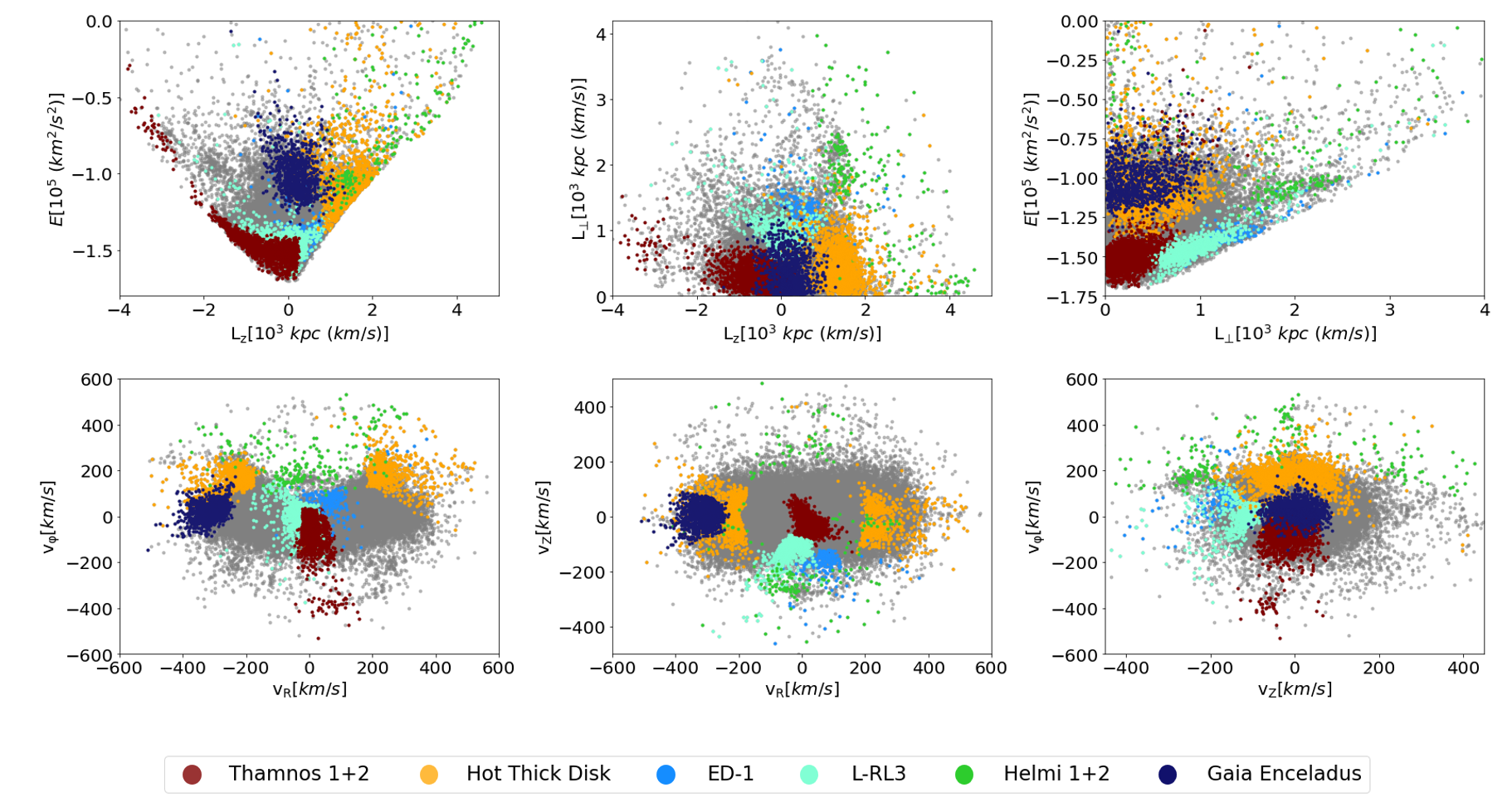}
  \caption{Substructures of the solar neighborhood halo identified by the GS$^{3}$ Hunter, where the different colors indicate different groups or clusters. The background in gray shows the stars that are not clustered in the method. The top panel shows the IOM energy-angular momentum space, and the bottom panel shows the velocity distributions, using the same colors for each component. As seen, 6 well-known substructures are revealed here.}
  \label{sun_ssss}
\end{figure*}

\section{Results} 
\label{section:result}
\subsection{Local halo results}

We apply the above algorithm to the two datasets respectively, primarily by inputting six-dimensional (6D) phase space parameters to the network, and then obtaining the groups or substructure candidates. When we have candidates, we first check whether they are reported in other literature. Those that are left over we assign as new candidates to be confirmed in the future. We have described the criteria by which we assign a group to a substructure (which might correspond 1 or more groups) above.

We run GS$^{3}$ Hunter using the 6D parameters ({\tt\emph x}, {\tt\emph y}, {\tt\emph z}, {\tt\emph v$_x$}, {\tt\emph v$_y$}, {\tt\emph v$_z$}) from the local halo sample of 51, 671 Gaia RVS  stars, setting the initial parameter {\tt siam$_D$} as 3. GS$^{3}$ Hunter detects 38 groups, of which 13 groups are associated with known structure, 12 are corresponding to the 6 well-known structures from previous work (one structure might correspond to a few groups), that is: Gaia-Enceladus-Sausage \citep{2018Natur.563...85H, 2018MNRAS.478..611B, 2003ApJ...585L.125B}, Hot Thick Disk \citep{2019A&A...632A...4D, 2018Natur.563...85H}, L-RL3 \citep{2023A&A...670L...2D}, Thamnos \citep{2018ApJ...860L..11K}, Helmi streams \citep{2022arXiv220513810R}, ED-1 \citep{2023A&A...670L...2D}. More comparisons are shown in the discussion. Our results are shown in Figure \ref{sun_ssss}, which illustrates the clustering features in the {\tt\emph E}-{\tt\emph L} and velocity space. Different colours represent different previously known structures and the grey dots are background.

Gaia-Sausage-Enceladus (GSE) is colored by midnightblue in Figure \ref{sun_ssss}. It is the most important substructure in the local halo because it contributes much to the inner halo and some estimate that it accounts for $\sim$ 40 $\%$ of the halo in the solar neighborhood \citep{2020MNRAS.492.3631M, 2018MNRAS.478..611B}. GSE debris is a major contributor to the inner halo, and was slightly more massive than the Magellanic Clouds prior to infall also mentioned in \citet{2018Natur.563...85H}. Recently, \citet{2023ApJ...944..169D} also pointed out that the inner halo can not be primarily composed of debris from a single massive, ancient merger event like GSE; there are at least four components contributing to inner halo. In this work, we have found 1 group that is clearly related to this structure, which contains 2572 stars. Here, the reference which we have based on to identify this structure is \citet{2018Natur.563...85H} and \citet{2022A&A...665A..57L} (see their Extended Data Figure Captions and Section 5.2).

The Hot Thick Disk is shown in orange in Figure \ref{sun_ssss}. This structure is also dominant in halo-like stars close to the Sun, with a metal-rich [Fe/H] ${\textgreater}$ $-$0.7 dex population. Also known as the ``Splash'', it is linked to the thick disk. This structure may have been born in the thick disk of the Milky Way before the massive accretion event that also changed their orbits \citep{2020MNRAS.494.3880B}. In our result, we find that 5 groups are related to this structure; to identify this structure we refer to \citet{2023A&A...670L...2D} (see their section 4.1).

The Thamnos structure is colored in maroon in Figure \ref{sun_ssss}. They were originally discovered by \citet{2019A&A...631L...9K} using a sample of RVS data from Gaia DR2; stars were identified with {\tt\emph v$_\varphi$} $\sim -200$ km s$^{-1}$ and {\tt\emph v$_\varphi$} $\sim -150$ km s$^{-1}$, respectively, as Thamnos1 and 2. Stars in Thamnos are in a low-inclination, mildly eccentric retrograde orbit. \citet{2023MNRAS.520.5671H} summarizes the chemical characterization of Thamnos. The distribution of Thamnos in the chemical plane is observed to be similar to that of the accretion population of low-mass satellite galaxies and the Milky Way. It is also noted that the abundance pattern of Thamnos is relatively unique and not aligned with any of the other substructures they have studied. In this work, we find 2 groups that are related to this feature, as defined by \citet{2019A&A...631L...9K} (see their Section 4).

% The Helmi stream is shown in lime green in Figure \ref{sun_ssss}. \textcolor{red}{We found 2 groups related to this structure, as defined by \citet{2022A&A...659A..61D}.} We also cross-matched with the data from Gaia DR3 \citet{2022arXiv220607937K} to obtain chemical and kinematic information for 158 stars. The middle and right panels of Figure \ref{helmi_stream} show the Kiel diagram and metallicity distribution functions (from the Gaia DR3 survey in \citet{2022arXiv220607937K}) for this structure with 2 groups. Through \citet{2022A&A...665A..46M} analysis, Helmi stream stars were found to have lower elemental abundances of Na, Mg, and Ca compared to other halo stars, including stars accreting from Gaia Enceladus. At [Fe/H] $\lesssim$ -1.5 dex, the difference in chemical abundances between Helmi stream and Gaia-Enceladus stars is significant, while in the region of high metallicity, some Helmi stream stars have the same elemental abundance ratios as with Gaia-Enceladus stars.

The Helmi stream is shown in lime green in Figure \ref{sun_ssss}. We found 2 groups related to this structure, as defined by \citet{2022A&A...659A..61D}. Through \citet{2022A&A...665A..46M} analysis, Helmi stream stars were found to have lower elemental abundances of Na, Mg, and Ca compared to other halo stars, including stars accreting from Gaia Enceladus. At [Fe/H] $\lesssim$ -1.5 dex, the difference in chemical abundances between the Helmi stream and Gaia-Enceladus stars is significant, while in the region of high metallicity, some Helmi stream stars have the same elemental abundance ratios as Gaia-Enceladus stars.

L-RL3 is shown in aquamarine in Figure \ref{sun_ssss}. \citet{2023A&A...670L...2D} discovered that this group is very similar to Cluster 3 identified in \citet{2022A&A...665A..57L, 2022A&A...665A..58R} and they, therefore, refer to it as L-RL3. L-RL3 contains a mixed population of stars. While most of the stars fit the profile of thick disk stars, a metal-poor fraction is also present. At the same time \citet{2022A&A...665A..58R} found no correlation between it and Kraken/Heracles by comparing their energy and angular momentum distributions. We continue to use this term to find 1 group that is related to this structure.

We also find the structure ED-1 shown in dodgerblue in Figure \ref{sun_ssss}, which was first detected in \citet{2023A&A...670L...2D} and most of stars in this structure are metal-poor, we have found 1 group are related to this structure.

In addition, we find GSH-11 (cluster \#11) is very similar to cluster \#38 shown in \citet{2022A&A...665A..58R}. GSH-11 is loosely distributed in the energy vs. angular momentum space in Appendix Figure \ref{new_gaia_11}, about 1.5 kpc from the Sun. This structure has a significant velocity distribution in both {\tt\emph x} and {\tt\emph z} directions in three-dimentional space, while most of the velocity values in the {\tt\emph y}-direction are around 0 km s$^{-1}$.

Note that we compare our results with some previous literature mentioned above, but different groups might have different structure names and locations using different methods and dataset; one possibility is that they might have detected different parts of the same structure. For example, in the discussion, we compare the results with the work using GALAH DR3 dataset (Figure \ref{E-Lz(Hist)}).

\begin{figure}[!t]
  \centering
  \includegraphics[width=0.45\textwidth]{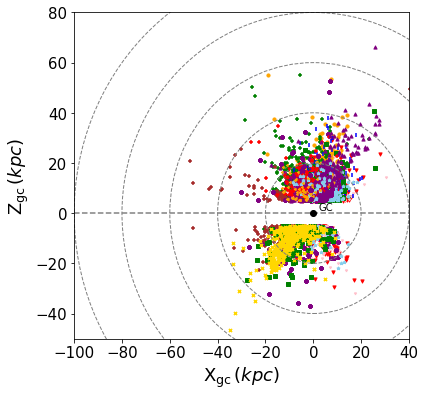}
  \caption{{\tt\emph x}-{\tt\emph z} distribution of the 29 candidate groups identified from the LAMOST halo K giants. The dashed line is the {\tt\emph r$_{gc}$} per 20 kpc. We use different colors and symbols to represent the different groups. One structure can has a few clustering groups.}
  \label{clusters_XZ_plot}
\end{figure}

\begin{figure}[!t]
  \centering
  \includegraphics[width=0.45\textwidth]{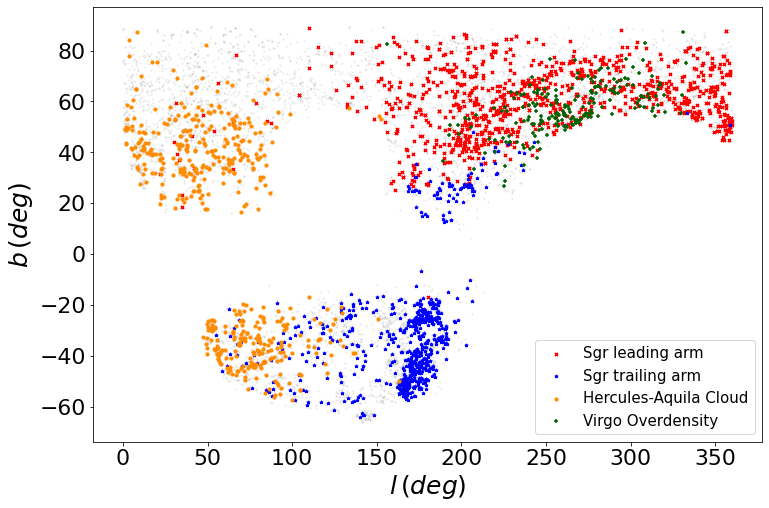}
  \caption{Sky coverage of the three structures identified from the LAMOST halo K-giant$:$ Sgr-leading part (red), Sgr-trailing part (blue), Hercules-Aquila Cloud (darkorange), Virgo Overdensity (darkgreen). The background (gray dots) is the sky coverage of the K-giants sample.}
  \label{clusters_lb_plot}
\end{figure} 

\begin{figure*}[!ht]
  \centering
  \includegraphics[width=0.9\textwidth]{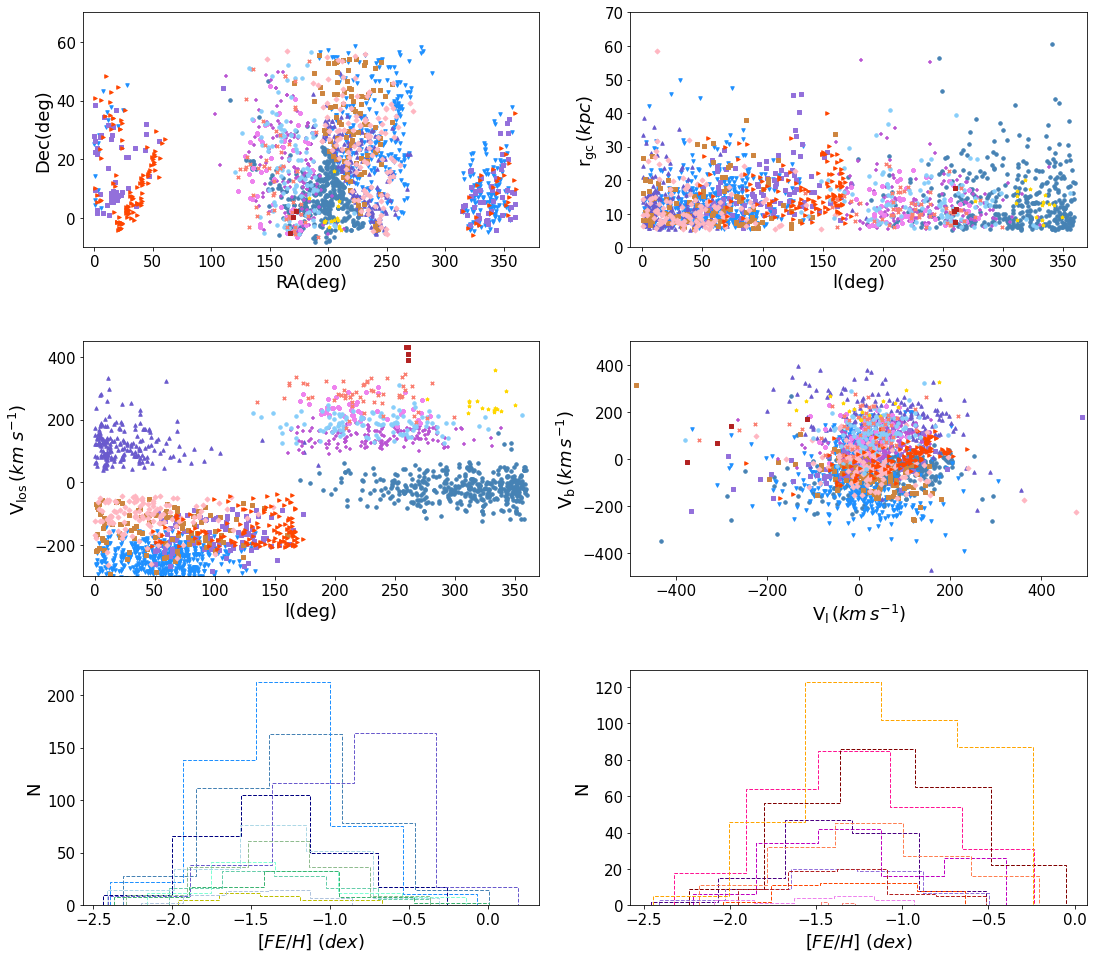}
  \caption{The figure shows the position and velocity distributions of the 13 unknown groups in the inner halo, as well as their metallicity distributions for the inner halo region.}
  \label{clusters_13_plot}
\end{figure*}

\begin{figure*}[!t]
  \centering
  \includegraphics[width=0.95\textwidth]{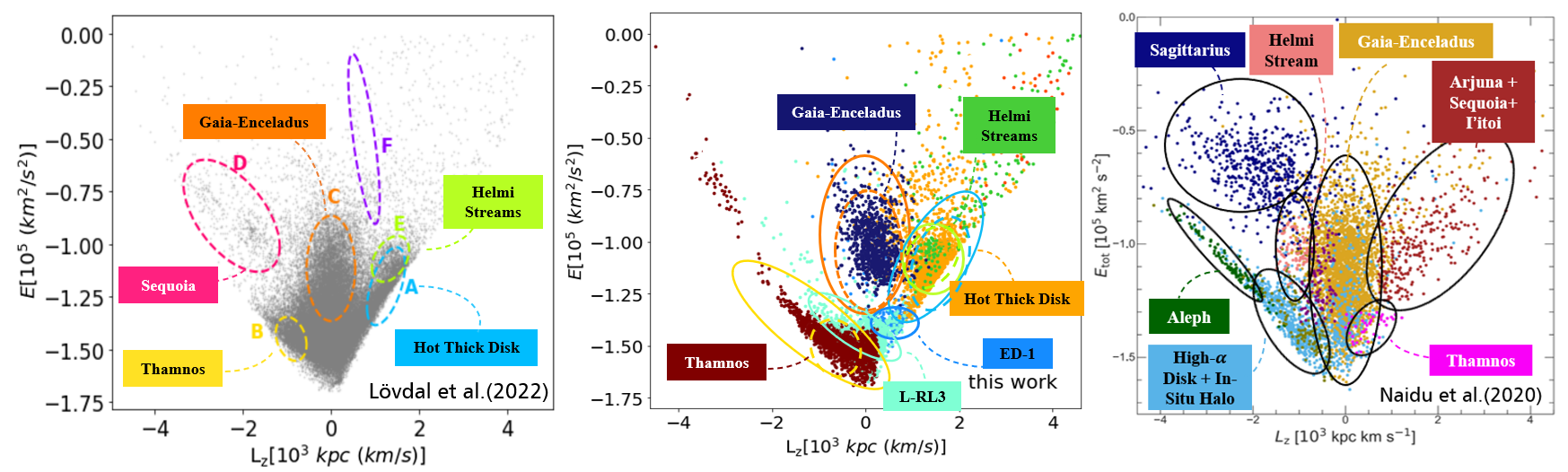}
  \caption{The left panel shows the positions of the six main groups in the {\tt\emph E}-{\tt\emph L$_z$} plane obtained by \citet{2022A&A...665A..57L}. Here we have identified the positions by circles, while different labels correspond to different structures. The middle panel shows the results of our work, where the structures represented by the different colors are shown by the name labels. Also we have circled the main part of the result here with a solid line using corresponding color by hand, while the dashed part is based on the left figure region \citet{2022A&A...665A..57L} but here we use 3$\sigma$ region to mark our substructure region. Note that our results are more extended. The right panel shows the results from \citet{2020ApJ...901...48N} obtained using H3 data; different colors correspond to different groups identified by ellipses. As seen, due to the different sample sky coverage and method, the left two have some differences with the right panel, but still have some of the same structures. Part of substructures in this work are more extended than the left panel such as the Hot Thick disk. The possible drawback in the purely kinematic space is not detecting the Sequoia clearly (see below Figure \ref{Sun(ELz)}).}
  \label{comparison}
\end{figure*}

\begin{figure*}[!t]
  \centering
  \includegraphics[width=0.95\textwidth]{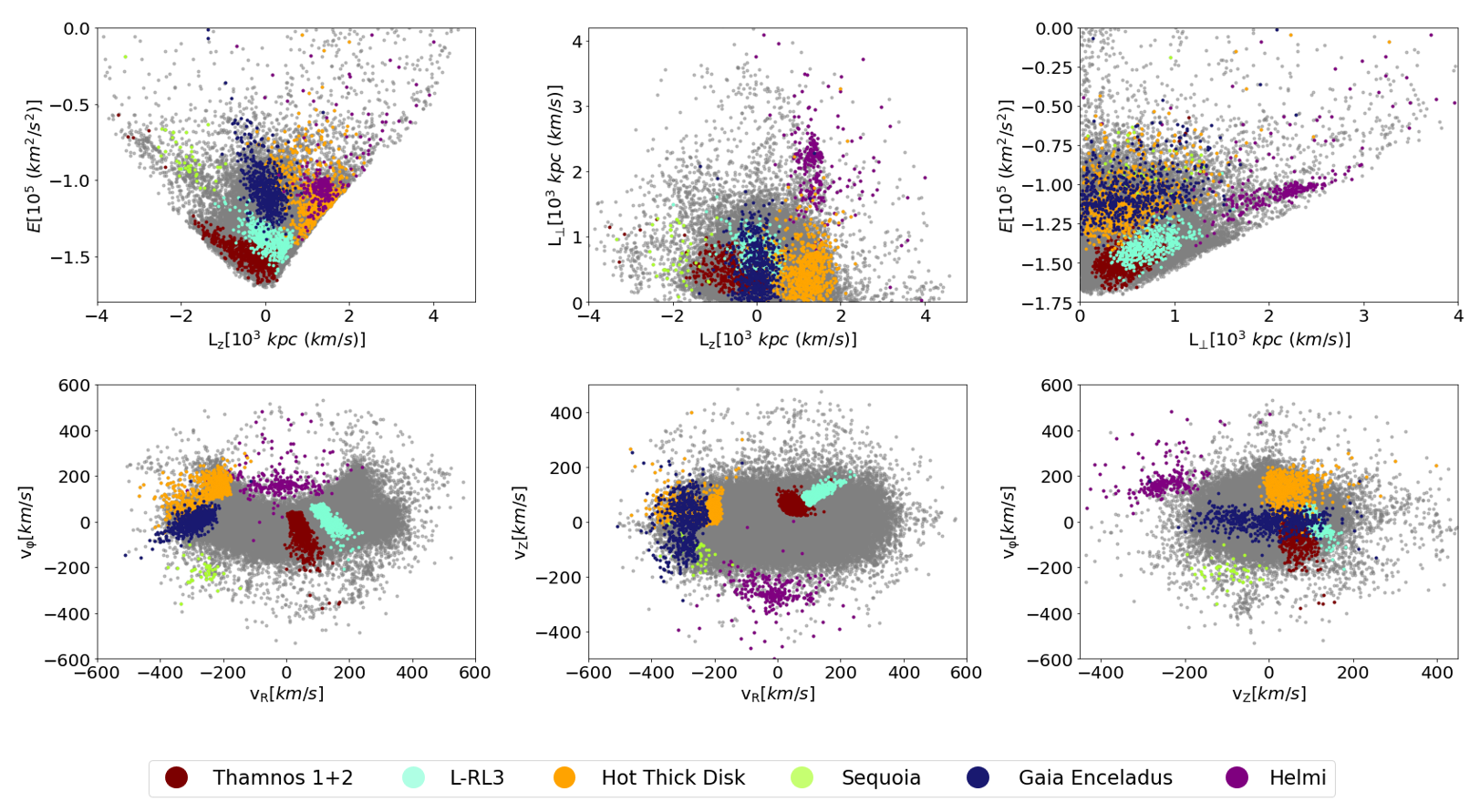}
  \caption{Substructures of the solar neighborhood halo identified by the GS$^{3}$ Hunter, which is similar to Figure \ref{sun_ssss}, but here Sequoia is detected.}
  \label{Sun(ELz)}
\end{figure*}

\begin{figure*}[!ht]
  \centering
  \includegraphics[width=6in]{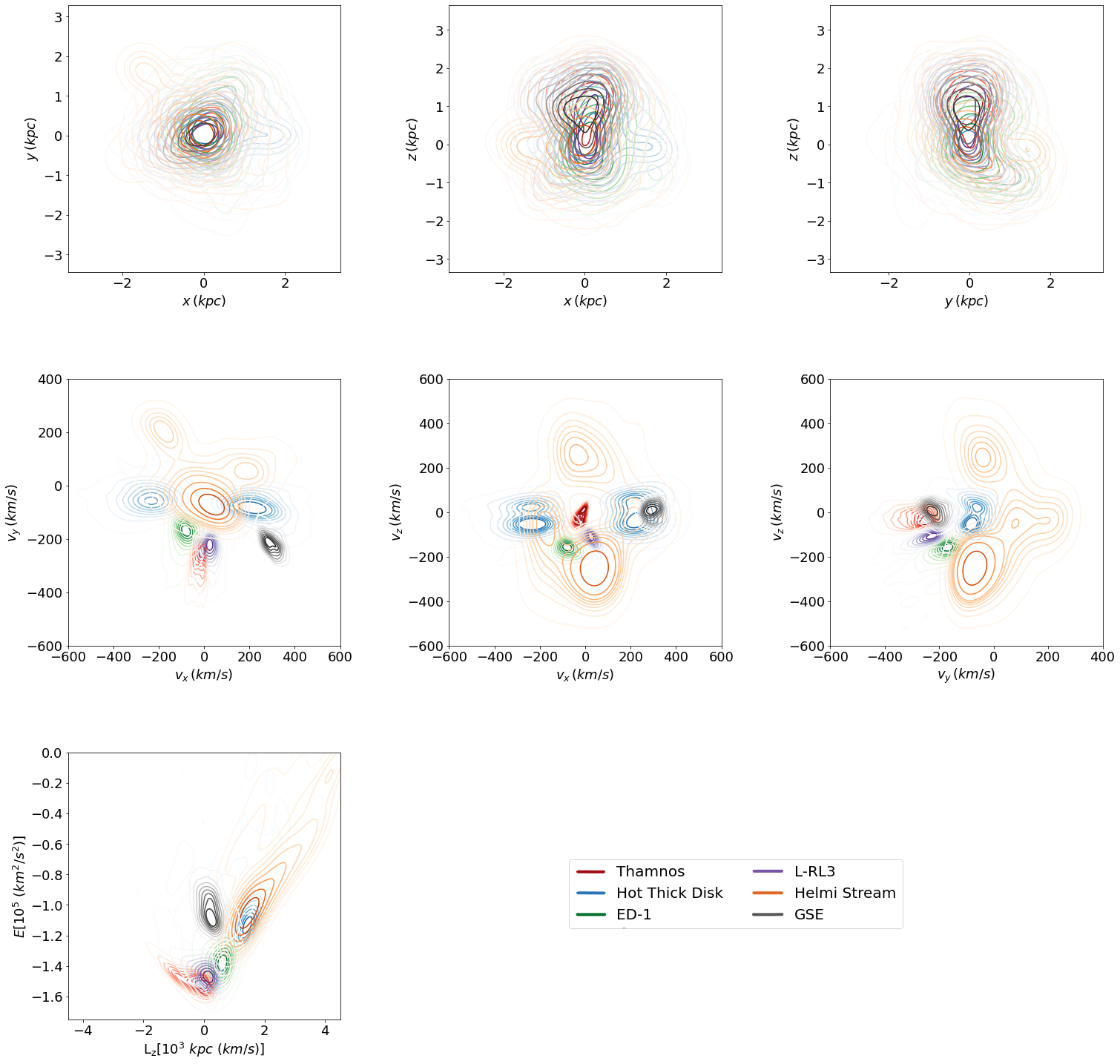}
  \caption{The first and second row present the density distribution of the result from GS$^{3}$ Hunter in spatial and velocity space, respectively, and the third row represents the result in IoM space.}
  \label{result(near the sun)_kde}
\end{figure*}

\begin{figure*}[!t]
  \centering
  \includegraphics[width=0.9\textwidth]{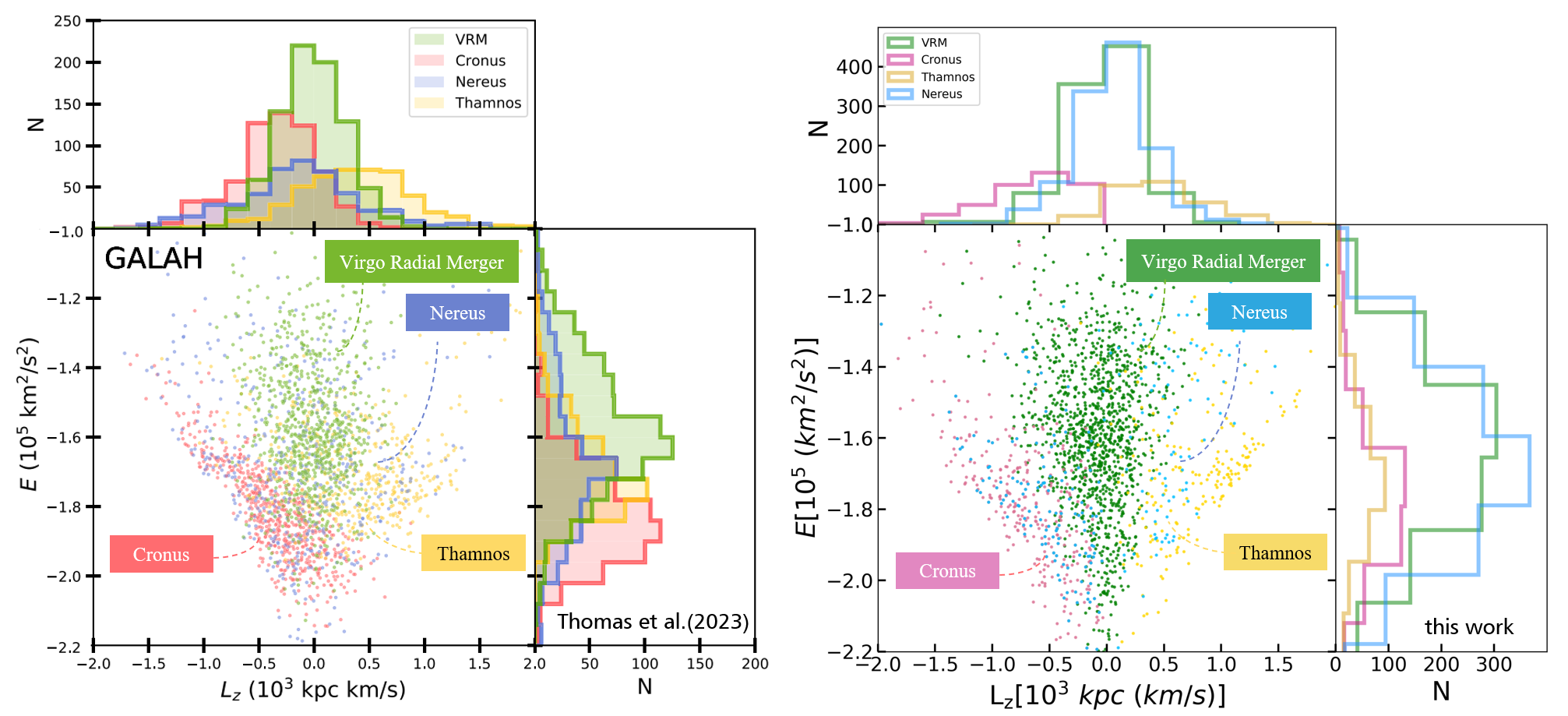}
  \caption{The comparison of our results of this method to \citet{2023ApJ...944..169D}. The left panel is from \citet{2023ApJ...944..169D} directly in the GALAH DR3 data part of Fig. 9, the different components are shown in green (VRM), red (Cronus), blue (Nereus), and yellow (Thamnos), respectively, and here we label them with different colored labels. In the right panel we show the results obtained by GS$^{3}$ Hunter using similar dataset, with the different components represented in green (VRM), purple (Cronus), blue (Nereus) and yellow (Thamnos). It can be seen that there is a good agreement between the two results although it is not the exactly same. We notice there might have some controversial analysis in different groups but here we only focus on the substructure or over-density hunting, moreover we currently speculate different groups have detected different parts of the same structure or stream using different methods and datasets, then they have different names for the structure in the halo.}
  \label{E-Lz(Hist)}
\end{figure*}

\subsection{Inner halo results}

Although we do not have enough sampling and careful fine tuning and detailed data process for inner halo larger region, as a by product and quick test, here we input the 6D information ({\tt\emph l}, {\tt\emph b}, {\tt\emph d}, {\tt\emph v$_r$}, {\tt\emph v$_l$}, {\tt\emph v$_b$}) from the selected 8099 LAMOST DR5 K giants in the halo to the deep learning network, with the initial parameter {\tt siam$_D$} set to 4.

Here we show the 29 candidate groups detected in the {\tt\emph x}-{\tt\emph z} plane, see Figure \ref{clusters_XZ_plot}. The sky coverage of the three well-known structures identified in the work are also shown in Figure \ref{clusters_lb_plot}, as seen, these three structures have some overlap. Meanwhile, Figure \ref{clusters_13_plot} shows the position and velocity distributions of the 13 unknown groups discovered in the inner halo, as well as their metallicity distributions for the inner halo region.

We obtained a total of 45 groups and find 29 candidates, 16 of these are related to known structures for the moment: Sagittarius stream\citep{2001ApJ...547L.133I, 2003ApJ...599.1082M}, we find 4 groups related to this structure; Virgo Overdensity \citep{2002ApJ...569..245N, 2012ApJ...753..145C, 2006ApJ...636L..97D}, we find 4 groups related to this structure; and Hercules–Aquila Cloud \citep{2007ApJ...657L..89B}, we find 2 groups related to this structure; and others in the previous literature. 

We have also analyzed Sagittarius stream (Sgr), virgo overdensity (VOD) and Hercules-Aquila cloud (HAC) for chemo-dynamical patterns obtained from GS$^{3}$ Hunter shown in the Appendix (see Fig. \ref{Sgr_more}, Fig. \ref{VOD} and Fig. \ref{HAC}), which are the properties of these two structures but we are not plannining to discuss the origins in this work. 

Meanwhile, we also find some structures that seem to have been identified in the previous work, and here we observe the characteristics of these structures. The cluster $\#$14 showed in Figure \ref{lamost_14} might be C-22 mentioned in \citet{2022MNRAS.516.5331M}; The  cluster $\#$15 showed in Figure \ref{lamost_15} might be the Slidr mentioned in \citet{2021ApJ...914..123I}; The cluster $\#$22 showed in Figure \ref{lamost_22} might be the combination of Slidr and Sylgr (see \citet{2021ApJ...914..123I}); The cluster $\#$26 showed in Figure \ref{lamost_26} must be a part of Gaia-1 (see \citet{2022MNRAS.516.5331M}); The cluster $\#$33 showed in Figure \ref{lamost_33} contains C-1 (see \citet{2022MNRAS.516.5331M}); The cluster $\#$35 showed in Figure \ref{lamost_35} might be the combination of C-1 and Fjörm structures (see \citet{2021ApJ...914..123I}), 6 in total in the appendix for the inner halo part analysis. 

In addition, 13 groups (1891 K-giants) cannot be linked to any known substructure (see the Figure \ref{clusters_13_plot}) as far as we know, so they may be related to unknown new substructures which will be investigated in more detail in our second work about the tomography of the Milky Way halo streams and substructure from Galactic centre 5 to 30 or 30 to 120\,kpc. The inner halo is not the key target for this work, and we aim to refine our method for the inner halo in future studies.

\section{Discussion} 
\label{section:discussion}
\subsection{Local halo substructures}
\label{section:disk part}

In this part, we show the 6 substructures corresponding to the 12 groups identified with GS$^{3}$ Hunter and compare them with the results in \citet{2020ApJ...901...48N} and \citet{2022A&A...665A..57L} for Gaia-Enceladus, Sequoia and Thamnos, etc. The phase space properties of these candidates may represent either a single structure (e.g., GSE) or several components of a structure. It's important to note that in our current work, it's normal for one structure or stream to correspond to one or more groups.

As shown in Figure \ref{comparison}, we recover clustering features associated with these substructure in the similar regions, in particular \citet{2022A&A...665A..57L}. We have approximately marked the locations of these substructures with different color labels in the Figure. It seems that Thamnos, Helmi and GSE structures in the energy-angular momentum space are more extended in our work, which does not come as a surprise, since it might mainly be due to the differences in the clustering methods different groups adopt. The distribution of these substructures in chemo-dynamical space was revealed and analyzed by \citet{2023A&A...670L...2D}. These substructures or streams found here will be worthwhile for further studying their internal properties, such as their specific populations, metallicity, and chemical abundances respectively in the future works.

The main substructures we found are GSE, Helmi stream, Hot Thick Disk, Thamnos, ED-1, L-RL3, but we did not find the Sequoia structure in the purely kinematics space \citep{2019MNRAS.488.1235M} with our method, compared to the results of \citet{2022A&A...665A..57L} and \citet{2020ApJ...901...48N}. This may be due to the fact that the structure is relatively dispersed in the six-dimensional kinematic space and is not well identified by our method or it might be mixed with other structures. We speculate that after adding the energy or other IOM parameter, we might recover this main structure, test for which shows in Figure \ref{Sun(ELz)} by adding the {\tt\emph E} to our algorithm, we have seen the famous structure. One may notice that our method, at this stage, can only be applied well to small batch such like a few kpc region. When it is applied to the large area such as inner halo region, it needs more patches or regions to search. This drawback will be solved step by step, but it will not affect much our conclusions focusing on the 6D parameters for this current work. To improve our methodology, we propose the following steps: $a).$ Conduct a more detailed analysis of the specific phase space distribution of different structures in future work. $b).$ Enhance the flexibility of our method by removing the fixed threshold parameter, allowing the network to learn and train different thresholds for various cluster distributions. However, for the current study, using a fixed value is reasonable.

We also present the density distribution of the results integrated by GS$^{3}$ Hunter across different parameter spaces. Our findings indicate that structures are clearly mixed in the {\tt\emph xyz} spatial position space, whereas individual structures are more distinctly shown in the velocity space and the integrals of motion (IoM) space, as shown in Figure \ref{result(near the sun)_kde}. It shows that the choice of input parameter space significantly impacts on our results.

Note that here we mainly use the local sample to compare with the \citet{2022A&A...665A..57L}, so the structure Sagittarius streams, alpha, High $\alpha$ Disk + In Situ Halo at the right panel of Figure \ref{comparison} are not shown well, our sample is not distributed in this range. But it seems that the right two panels show that the GSE can be more extended than the left panel. At the same time, some of the structures we have found, such as Hot Thick Disk and Helmi stream, have a few stars that are relatively looser and not compact, which might imply these structures may have other extensions. In this work, we mainly present our new method, further analysis of the structure will be further explored in the subsequent work.

%Interestingly, when we add the E and L parameters to the network we find Sequoia is clearer after a test, it might be implying that the IOM might be better than pure kinematics for this method, however, the ED1 is not as clear as the current results, so as the second step, more results will be shown in the next work.

In the meantime, we also analyze the local halo component using the GALAH DR3 data. Our data selection steps are equivalent to those employed by \citet{2023ApJ...944..169D}.
We input the GALAH DR3 sample into GS$^3$ Hunter and identified a total of 30 clusters/groups. Comparing with \citet{2023ApJ...944..169D} for each structure in  Fig. \ref{E-Lz(Hist)}, it is noticed that our results are in good agreement.  Because the local stellar halo contains a variety of substructures as in Fig. \ref{E-Lz(Hist)}, different GSE stellar selection methods may choose different mixtures of these substructures; more details can also be found in \citet{2023arXiv231009376D, 2023AAS...24122802D, 2022ApJ...932L..16D}.

\subsection{Unknown structures properties of the local halo}
\label{section:Analysis of unknown structures}

In this work, we discover many unknown new structures (as far as we know) or substructures or streams in the local halo, and here we briefly analyze and introduce the characteristics of some of the unknown structures with high star fraction (at least 95\%) in the definition. The unknown structure properties can also be found in the Appendix of this paper. 

\subsubsection{GSH-8}

GSH-8 has a similar distribution in energy vs. angular momentum space as the structure Hot-thick-disk shown in Figure \ref{new_gaia_8}, and is more diffuse in the three-dimensional position space. The distribution of GSH-8 is more concentrated in the three-dimensional velocity space. We can note that the velocity values of the GSH-8 structure in the {\tt\emph y}-direction are mainly concentrated around $-$125 km s$^{-1}$ with the colour magnitude clustering features in the last panel. Based on the isochrone in the last H-R subplot, we can see that most of the stars in this structure have a log(age) of about 9.8.

\subsubsection{GSH-9}

GSH-9 is a group of 1924 stars shown in Figure \ref{new_gaia_9}, which has a relatively large number of stars and may contain some contaminations. This group is mainly in the low energy region (E $\sim$ [-1.8,-1.2] (10$^5$\ km$^2$\ s$^{-2}$)) and has a large scatter in the {\tt\emph v$_z$}-{\tt\emph v$_\phi$} region. In the last panel, the isochrone shows that most of the stars in this structure have a log(age) of about 9.5.

% \subsubsection{GSH-11}

% GSH-11 is loosely distributed in the energy vs. angular momentum space in figure \ref{new_gaia_11}, about 1.5 kpc from the Sun. This structure has a significant velocity distribution in both $x$ and $z$ directions in three-dimentional space, while most of the velocity values in the $y$-direction are around 0 km s$^{-1}$. \textcolor{red}{This new structure is very similar to cluster \#38 shown in \citet{2022A&A...665A..58R}.}

\subsubsection{GSH-14}

GSH-14 is a more relaxed structure in the energy-angular momentum space as shown in Figure \ref{new_gaia_14}. The continuous part in the velocity space is mostly larger in the direction of {\tt\emph v$_z$}, which ranges from 200 km s$^{-1}$ to 400 km s$^{-1}$. In the last panel, the isochrone shows that most of the stars in this structure have a log(age) of about 9.2.

\subsubsection{GSH-16}

The group GSH-16, as shown in Figure \ref{new_gaia_16}, shows a certain overlap in energy-angular momentum space with the position of the new structure ED-1 found by \citet{2023A&A...670L...2D}, but it is less consistent in velocity space and GSH-16 covering a slightly larger region. As seen in the isochrone of the last H-R diagram, we can see that most of the stars in this structure have a log(age) of about 9.6.

\subsubsection{GSH-20}

GSH-20 seems to give a good linear structure showed in Figure \ref{new_gaia_20}, based on the narrow structure, the progenitor of the GSH-20 structure may be a globular cluster, which need further investigations. The component in the {\tt\emph z}-direction of the galactic plane and the {\tt\emph y}-direction of velocity space is about 0 km s$^{-1}$, while the distribution of velocity direction in {\tt\emph x} and {\tt\emph z} is about 200 km s$^{-1}$ to 500 km s$^{-1}$. In the last panel, when the isochrone is taken into account, it can be observed that the age of this structure have a log(age) of about 8.05. 

\subsubsection{GSH-27}
The structure GSH-27 in Figure \ref{new_gaia_27} is in the low-energy space and is relatively dispersed. At the same time the distributions of GSH-27 and Thamnos in the energy angular momentum space overlap each other. When the isochrone is taken into account in the last panel, it can be observed that the age of this structure have a log(age) of about 9.7.

\subsubsection{GSH-30}
In Figure \ref{new_gaia_30}, this structure is in the low-energy space region, and it can be seen that GSH-30 also overlaps each other with the Thamnos structure. However, the distribution of the Thamnos structure in the velocity space is more widely distributed, mainly in {\tt\emph v$_R$}-{\tt\emph v$_\phi$} and {\tt\emph v$_R$}-{\tt\emph v$_z$}. The distribution of this structure, GSH-30, in the velocity space is more concentrated. When the isochrone is taken into account in the last H-R diagram, it can be observed that the age of this structure have a log(age) of about 9.6.

\subsubsection{GSH-36}
In Figure \ref{new_gaia_36}, we show the propeites of the cluster GSH-36. This cluster has a more diffuse, stream-like distribution in the energy angular momentum space, which needs more analysis. It is also observed that the structure is 500 km s$^{-1}$ in {\tt\emph v$_\phi$} and a larger distribution in the {\tt\emph v$_z$} direction from $-$400 km s$^{-1}$ to 200 km s$^{-1}$. The isochrones shows that most of the stars in this structure have a log(age) of about 7.5.

\subsection{Inner halo structures}
\label{section: inner halo part}

Our results for the LAMOST DR5 K giants do not include the Monoceros Ring substructure because we exclude the stars with {\tt\emph z}$=5$ kpc. \citet{2003ApJ...588..824Y} also have pointed out that the structure is mainly located in the $-$20$^{\circ}$ to 30$^{\circ}$ disk latitude range so we think the main reason is that we do not cover the Monoceros Ring range in our data selection, and the origin of which might be from from the disk flaring \citep{wang2018b}. By the analysis of the dynamical and chemical abundance distributions of the VOD and HAC member stars, \citet{2023A&A...674A..78Y} find that among the orbital parameters, the distributions of the eccentricity, the apocenter, and the maximum height from the Galactic disk show consistency. Both structures are located in low angular momentum and high energy orbits with similar spatial distribution of chemical abundances. Ultimately, they conclude that the two structures may have originated from the merger of the same dwarf galaxy. \citet{2019MNRAS.482..921S} utilized RR Lyrae stars to study the orbital properties of HAC and VOD, they revealed that a very similar spatial distribution of kinematics for HAC and VOD stars. From this they concluded that the two structures may be part of the same accretion event. We also have some observational analysis in the Appendix but defer the origin discussion to the future work.

\section{Conclusions}
\label{section:Conclusion}

In this work, we have developed a new method (in section \ref{section:method}) that we call Galactic-Seismology substructures and streams Hunter (GS$^{3}$ Hunter) to find the structures or substructures or streams of the Galactic solar neighborhood halo (local halo) and the Galactic inner halo. We first verify the feasibility of our algorithm with the FIRE suites of mock Milky Way-like simulated galaxies, which contains a total of 594,409 stars in the FIRE simulation data and each star is labeled with a real structure. After applying GS$^{3}$ Hunter, we successfully find the corresponding 8 real structures with different number of stars (in sec.\ref{section:method}).

Next, the data used in this work are Gaia EDR3 publicly available RVS sample and LAMOST DR5 K giants, we obtain the sample by analyzing the data (in sec.\ref{section:data}), finally 51671 stars from Gaia RVS sample (local halo) and 8099 stars from LAMOST DR5 K giants (Inner halo as by-product). After applying our method to the two datasets, we have detected a total of 38 clusters in the stellar local halo near the Sun, of which 13 are identified as associated with previously identified halo substructures. We find 1 group associated with GSE structure; 5 groups associated with hot thick disk structure; 1 group associated with L-RL3 structure; 2 groups associated with Thamnos structure; 2 groups associated with Helmi stream Structure\footnote{In addition, we would like to further analysis cluster 24. Due to the small size of this cluster and the loose distribution of a few stars, our calculated fraction of cluster 24 is only 80\%. But based on the location of most of the points of cluster 24, we have also included it as part of the Helmi stream.}; 1 groups linked to ED-1 structure. 1 cluster in this work is named as cluster \#11 which is corresponding to the cluster \#38 shown in \citet{2022A&A...665A..58R}.

Moreover, we also detect 45 clusters in the part of the inner halo ($|${\tt\emph z}$|$ ${\textgreater}$ 5 kpc). 16 groups within these cluster features are identified as being related to known substructures. We found 4 groups related to Sagittarius structure\footnote{It is worth mentioning that we have also grouped GSH-5 into the part of the leading arm of Sgr. Since some of the stars in this group we identified have a looser distribution, they did not pass our filters in data fraction value. However, by carefully analyzing the distribution of most of the stars in this cluster, we decide to classify it as part of the leading arm of Sgr.}; 4 groups related to virgo over-density structure; 2 groups related to Hercules-Aquila Cloud. 6 groups are mentioned in the recent literature. Furthermore, we also have analyzed some of these structures in more details with chemical and dynamical information, and we present the results in sec.\ref{section:result}.

%While our precursor investigation on the inner halo has led to some promising insights, as the focus of this optimisation of our method does not particularly focus on the inner halo, we will leave the detailed investigation to future studies.

% Note that we need to re-design to search the inner halo larger region in the next work, here we just have a quick investigation without fine tuning \footnote{So it is not strange at all that we do not find some more other streams, we defer an detailed investigation into the inner halo or outer halo to the second contribution.}, but the clustering features here is quite promising. 

More specifically, for the local halo, we find a total of 38 clusters, 21 of which we used as candidates (data fraction $>$ 95\%), and 13 of these clusters we selected to be associated with known structures. 8 clusters are currently considered as unknown structures and are presented above as our new discoveries. For the inner halo, we find a total of 45 clusters, 29 of which are considered as candidates (data fraction $>$ 95\%), and 16 of these clusters are selected to be related to known structures and 13 of these are unknown.

In discussion part of Sec. \ref{section:disk part}, we also have recovered the four substructures mentioned in \citet{2023ApJ...944..169D}, which is implying that different groups might have found different parts of the same stream with different methods and names, the inner halo or local halo will need more investigations in the future, for example, as mentioned in \citet{2022ApJ...939....2A} the fixed potential might give incorrect stream conclusion in action space. 

As a prospect, we will investigate more for the inner and outer halo region with action-angles and some other dynamical-chemical parameters as the second step, in the future we will apply our method to more Milky Way and local group streams research. 

\section*{Acknowledgements}

We would like to thank the anonymous referee for his/her very helpful and insightful comments. We are grateful for some comments or interactions from Heidi Newberg, Changjiang, etc. We acknowledge the National Key R \& D Program of China (Nos. 2021YFA1600401 and 2021YFA1600400).  TTG acknowledges  financial support from the Australian Research Council (ARC) through an Australian Laureate Fellowship awarded to Joss Bland-Hawthorn. Y.S.T. acknowledges financial support from the Australian Research Council through DECRA Fellowship DE220101520. L.Y.P is also supported by the National Natural Science Foundation of China (NSFC) under grant 12173028, the Sichuan Science and Technology Program (Grant No. 2020YFSY0034), the Sichuan Youth Science and Technology Innovation Research Team (Grant No. 21CXTD0038). The Guo Shou Jing Telescope (the Large Sky Area Multi-Object Firber Spectroscopic Telescope, LAMOST) is a National Major Scientific Project built by the Chinese Academy of Sciences. Funding for the project has been provided by the National Development and Reform Commission. LAMOST is operated and managed by National Astronomical Observatories, Chinese Academy of Sciences. This work has also made use of data from the European Space Agency (ESA) mission {\it Gaia} (\url{https://www.cosmos.esa.int/gaia}), processed by the {\it Gaia} Data Processing and Analysis Consortium (DPAC, \url{https://www.cosmos.esa.int/web/gaia/dpac/consortium}). Funding for the DPAC has been provided by national institutions, in particular the institutions participating in the {\it Gaia} Multilateral Agreement.

%{2018MNRAS.481.3442M} {2019ApJ...872..152I} {2022ApJ...930L...9M}

\appendix

\section{local halo unknown clustgers}
\begin{figure}[!h]%调节图片位置，h：浮动；t：顶部；b:底部；p：当前位置
  \centering
  \includegraphics[width=0.9\textwidth]{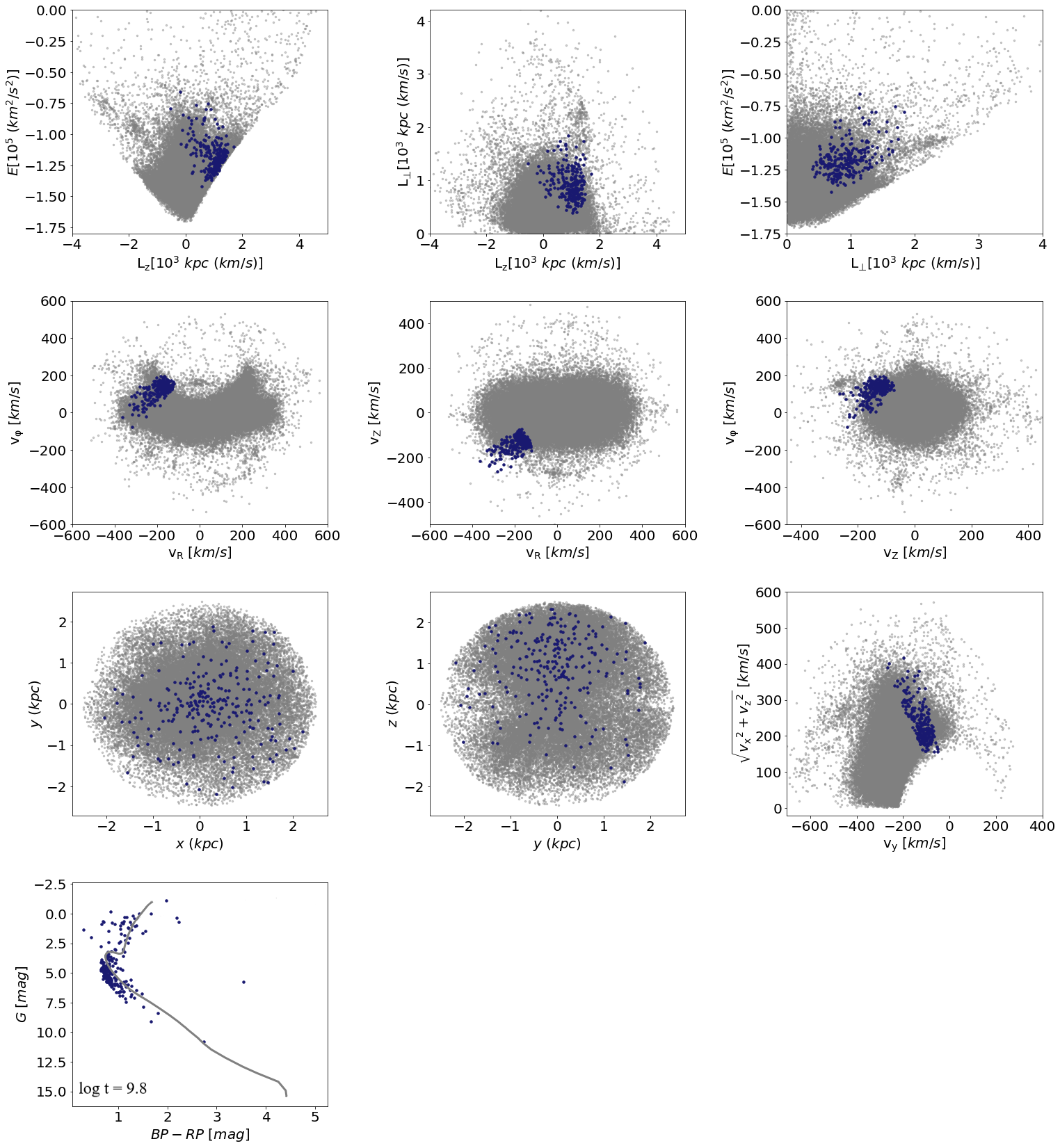}
  \caption{The three panels in the first row show the spatial distribution of the IoM for our unknown cluster $\#$8, while the second and third row present the 6D distribution of cluster $\#$8 in velocity and spatial space, respectively. For reference, the gray background shows all our halo samples. The last panel on the bottom left presents the CMD of cluster $\#$8, while we show its PARSEC isochrones \citep{2012MNRAS.427..127B} using a solid gray line, and the corresponding age information is also labelled in this panel.}
  \label{new_gaia_8}
\end{figure}

\begin{figure*}[!h]
  \centering
  \includegraphics[width=0.9\textwidth]{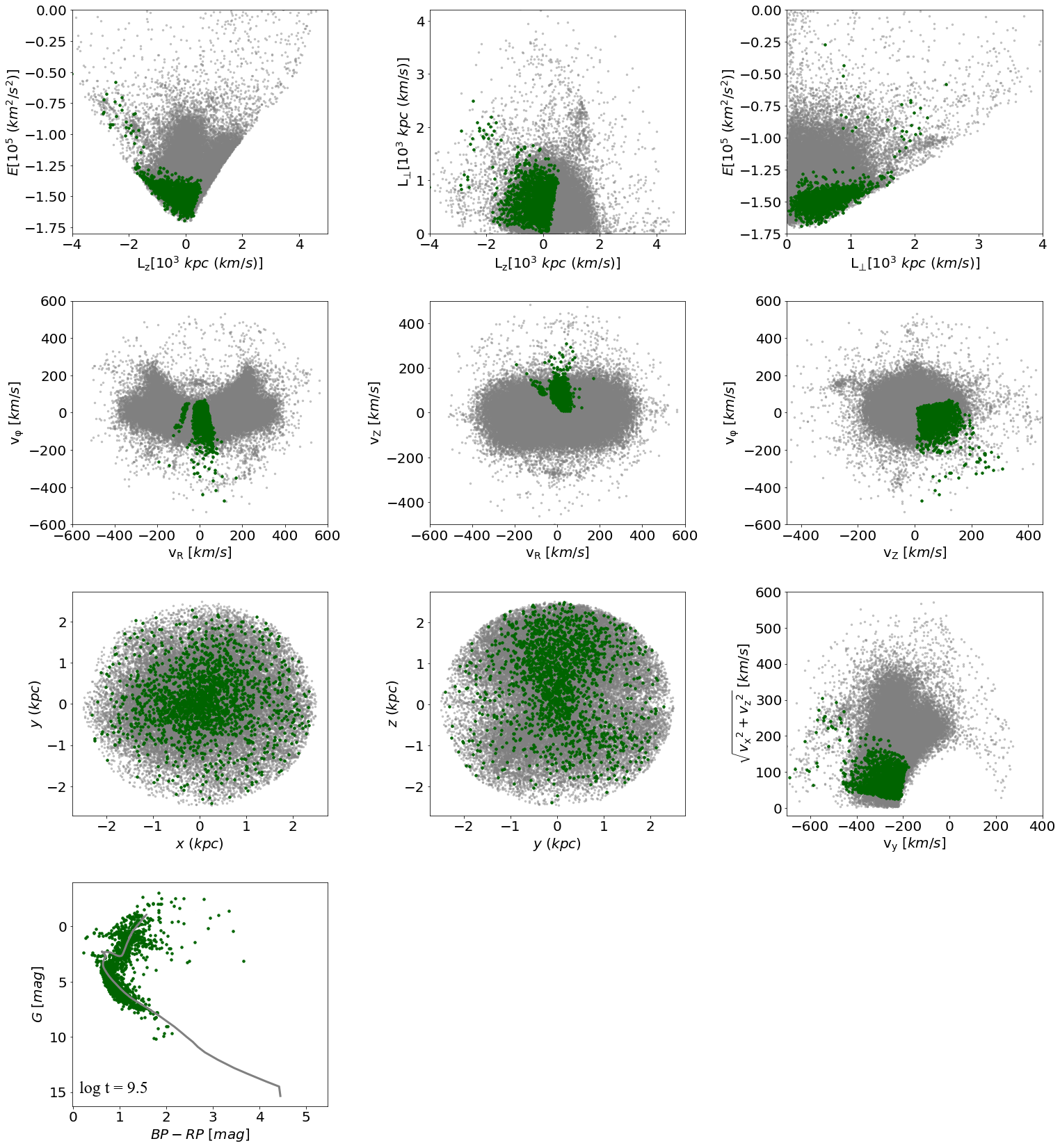}
   \caption{Cluster $\#$9 in IoM and 6D phase space, also the distribution of CMD. Similar to the Figure \ref{new_gaia_8}.}
  \label{new_gaia_9}
\end{figure*}

\begin{figure*}[!h]
  \centering
  \includegraphics[width=0.9\textwidth]{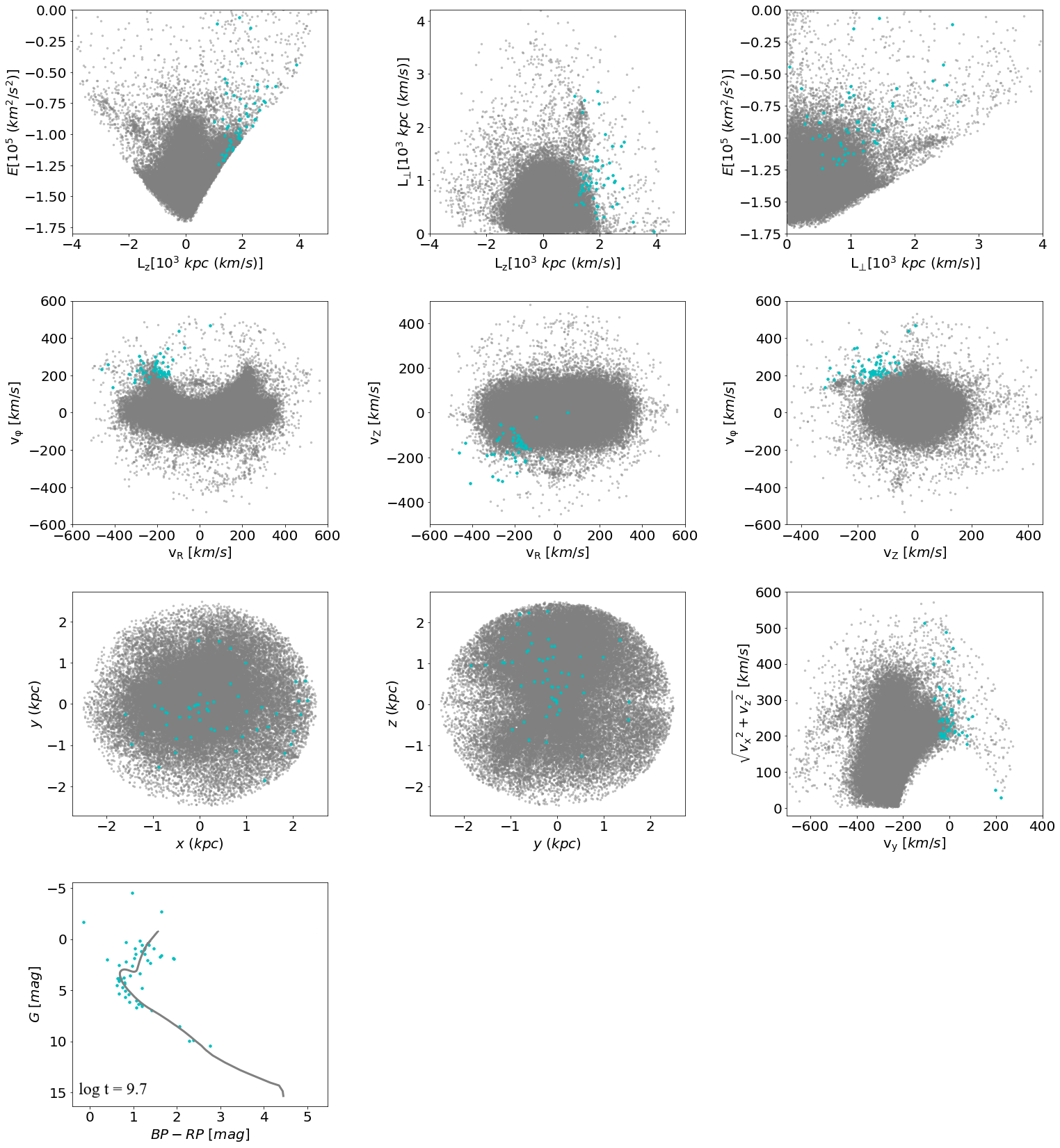}
  \caption{Cluster $\#$11. Similar to the Figure \ref{new-gaia-8}, which is corresponding to cluster \#38 mentioned in \citet{2022A&A...665A..58R}.}
  \label{new_gaia_11}
\end{figure*}

\begin{figure*}[!h]
  \centering
  \includegraphics[width=0.9\textwidth]{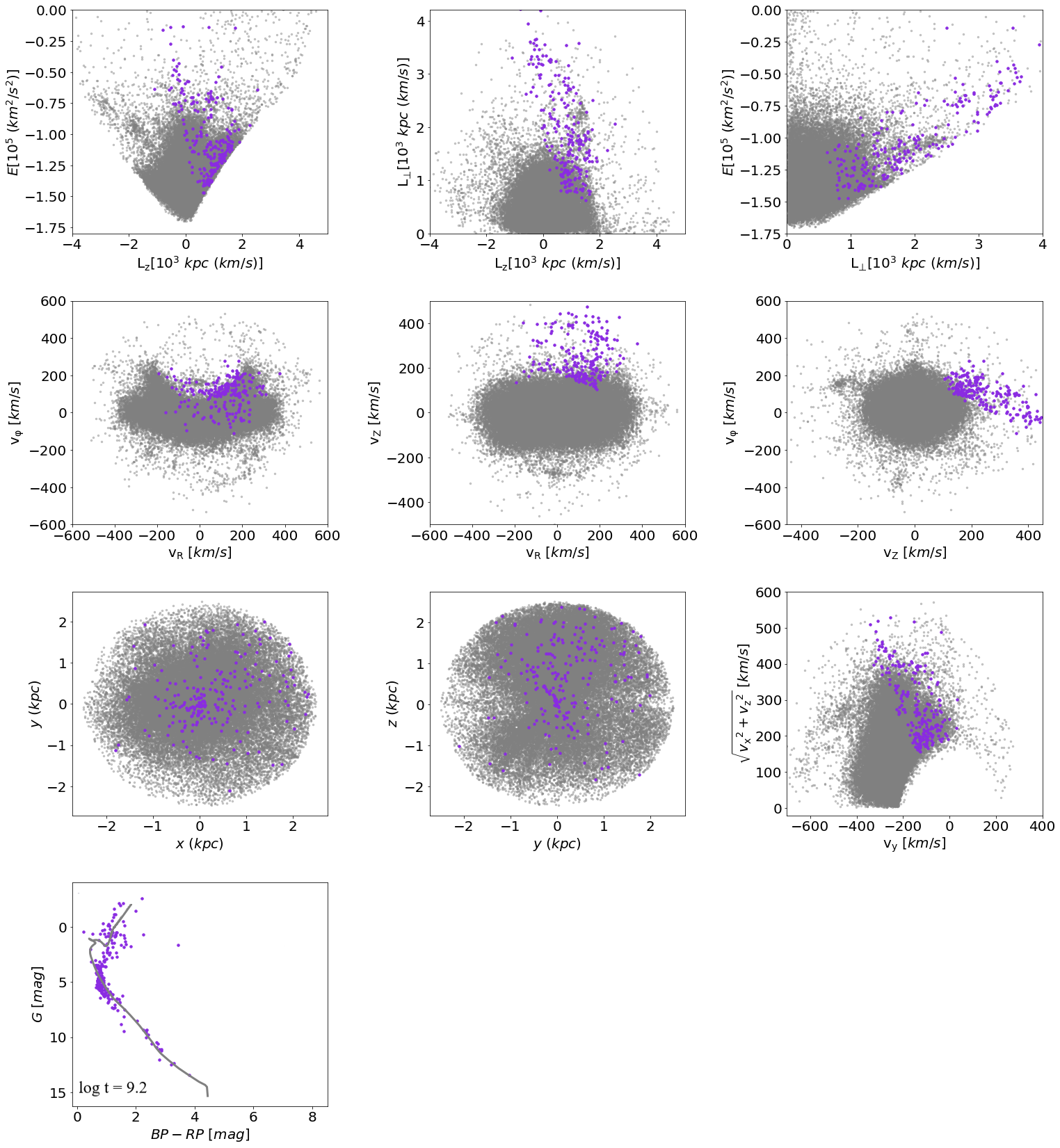}
  \caption{Cluster $\#$14. Similar to the Figure \ref{new_gaia_8}.}
  \label{new_gaia_14}
\end{figure*}

\begin{figure*}[!h]
  \centering
  \includegraphics[width=0.9\textwidth]{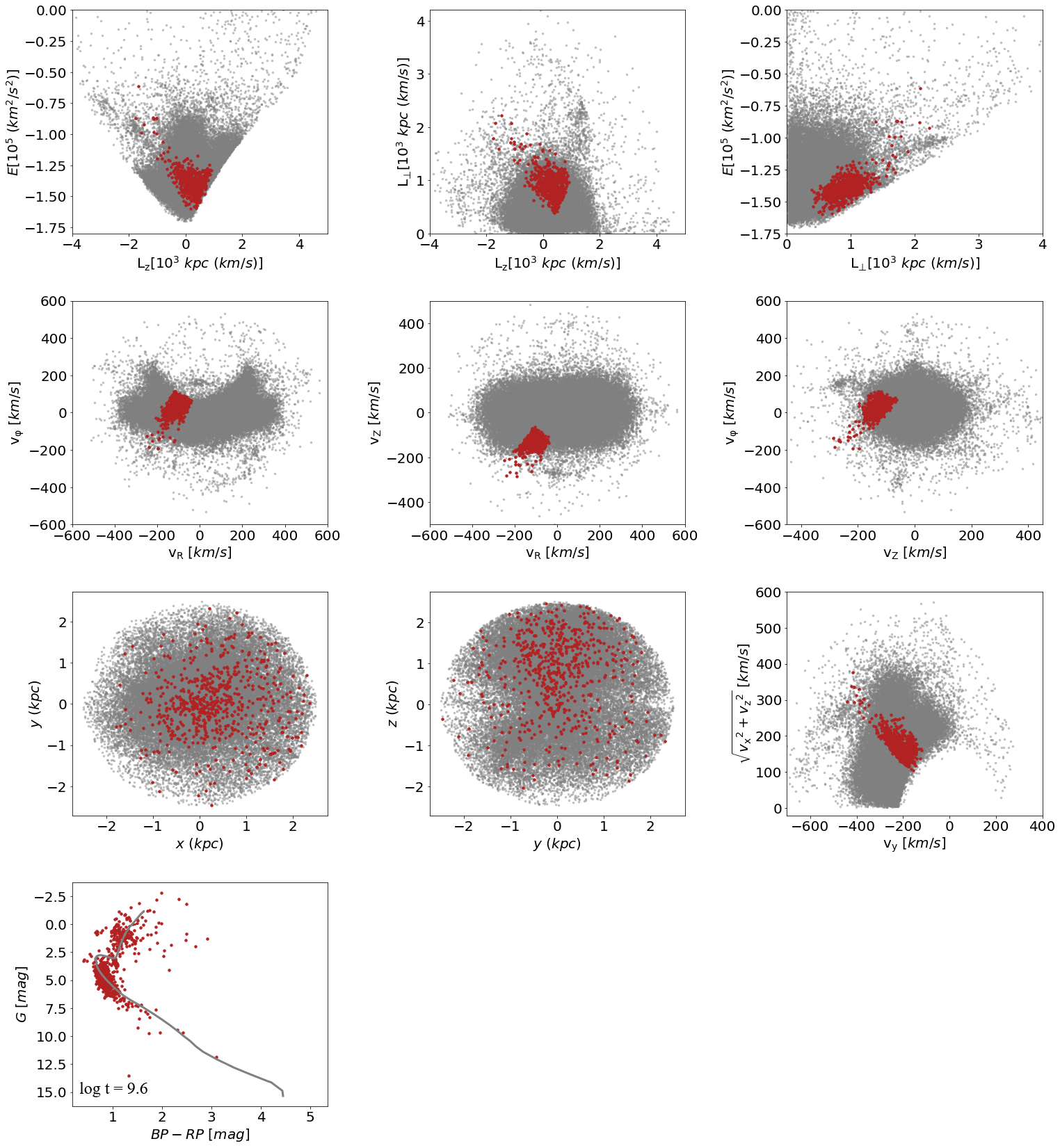}
  \caption{Cluster $\#$16. Similar to the Figure \ref{new_gaia_8}.}
  \label{new_gaia_16}
\end{figure*}

\begin{figure*}[!h]
  \centering
  \includegraphics[width=0.9\textwidth]{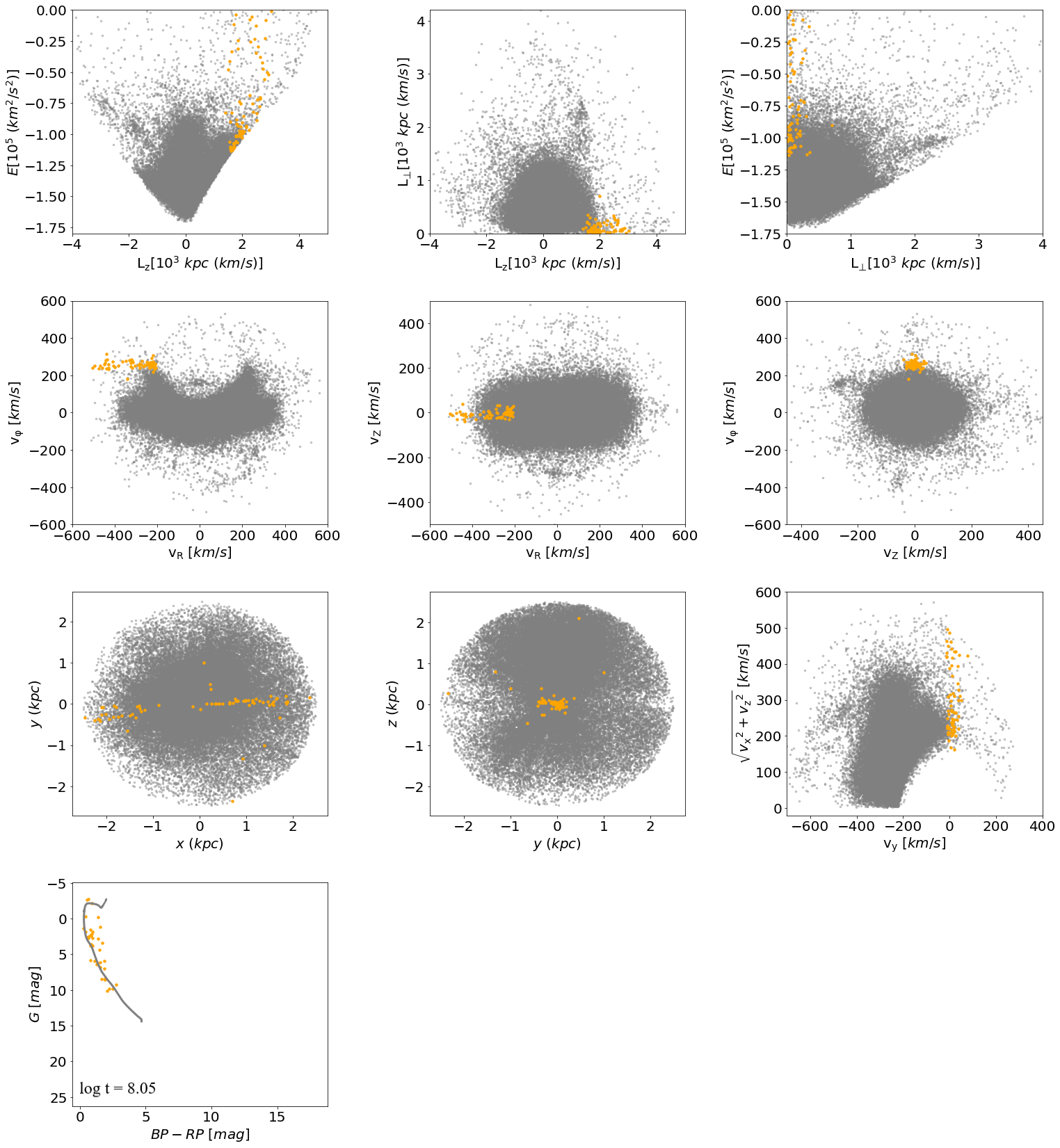}
  \caption{Cluster $\#$20. Similar to the Figure \ref{new-gaia-8}.}
  \label{new_gaia_20}
\end{figure*}

\begin{figure*}[!h]
  \centering
  \includegraphics[width=0.9\textwidth]{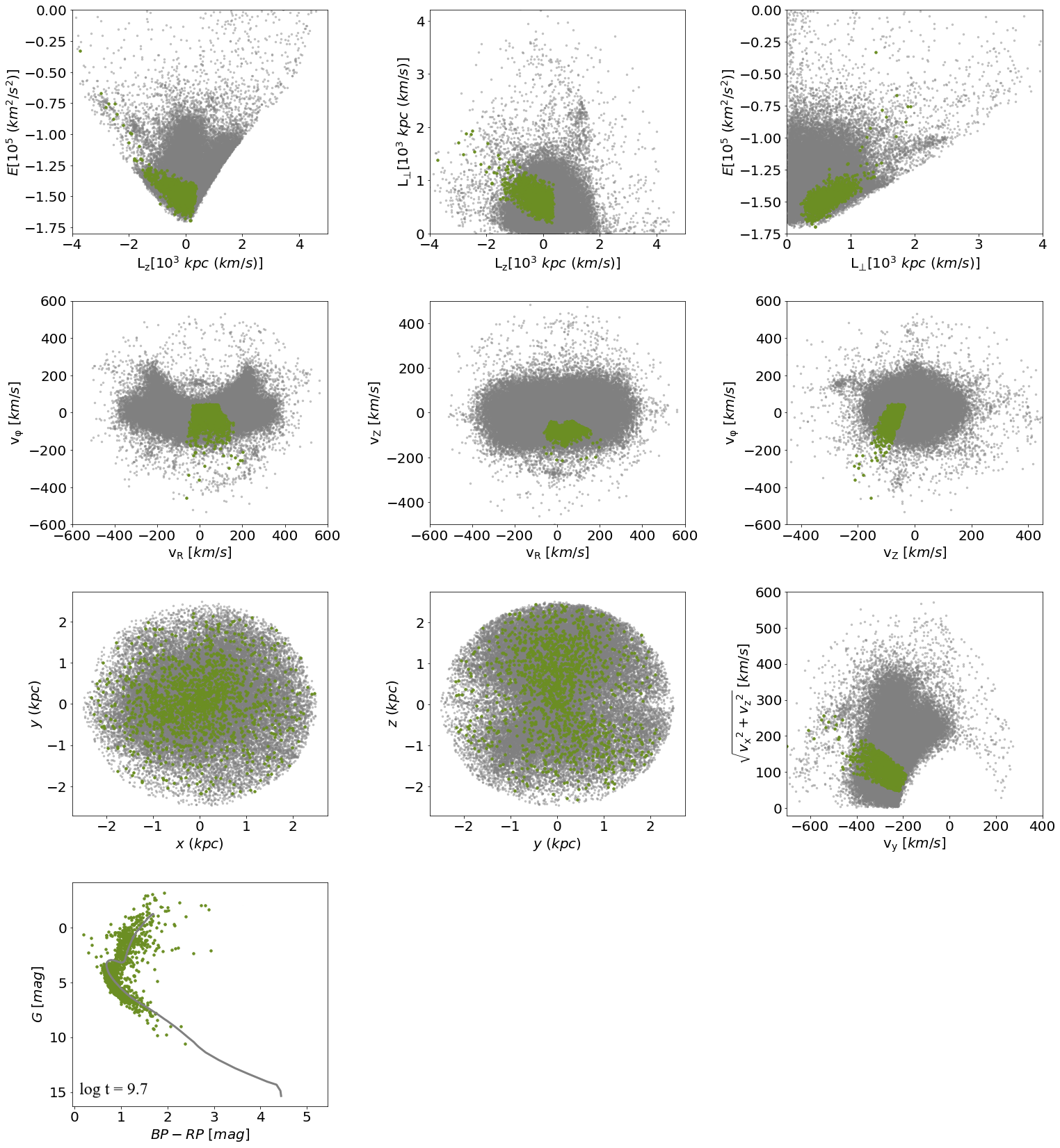}
  \caption{Cluster $\#$27. Similar to the Figure \ref{new-gaia-8}.}
  \label{new_gaia_27}
\end{figure*}

\begin{figure*}[!h]
  \centering
  \includegraphics[width=0.9\textwidth]{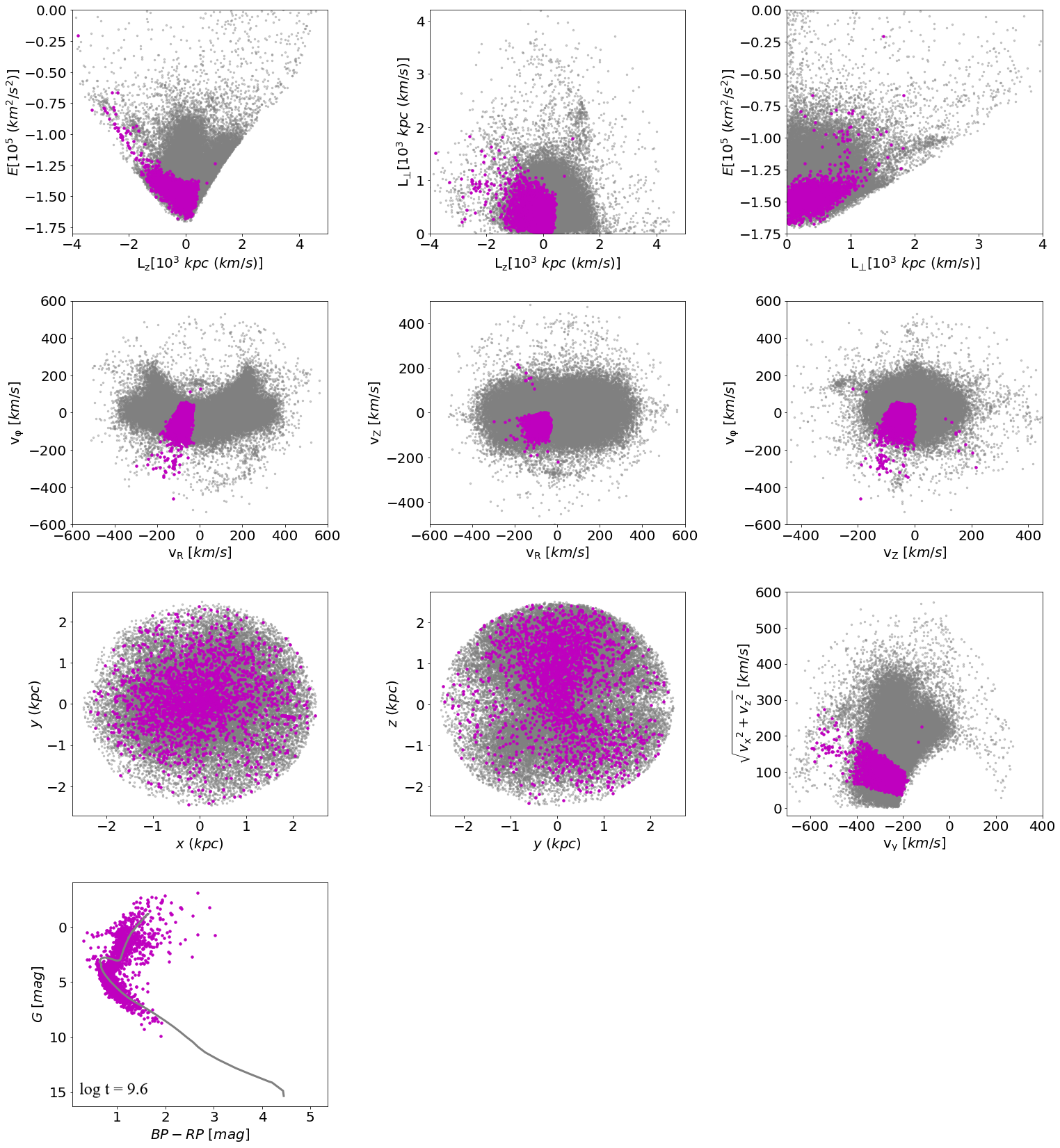}
  \caption{Cluster $\#$30. Similar to the Figure \ref{new-gaia-8}.}
  \label{new_gaia_30}
\end{figure*}

\begin{figure*}[!h]
  \centering
  \includegraphics[width=0.9\textwidth]{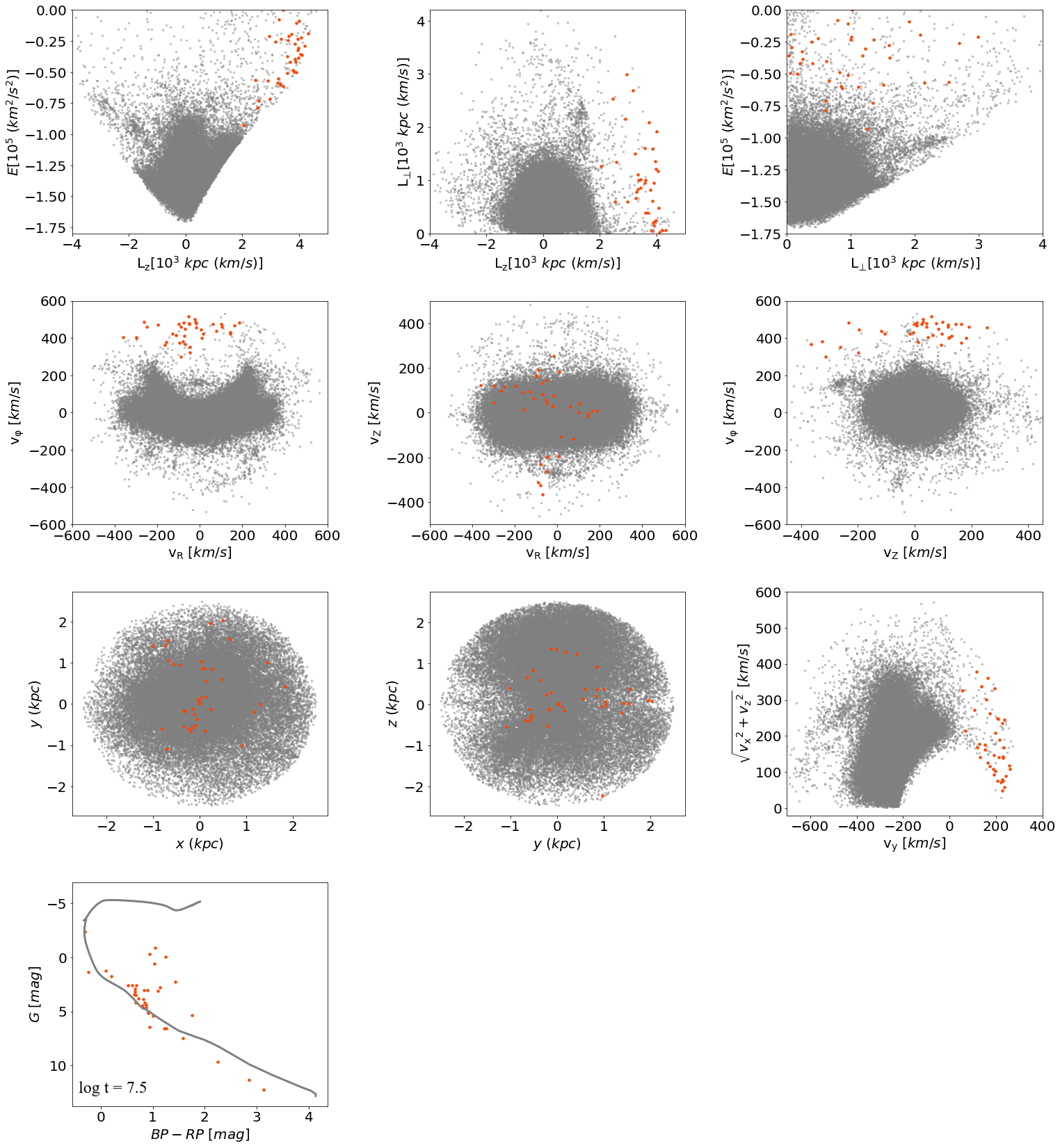}
  \caption{Cluster $\#$36. Similar to the Figure \ref{new-gaia-8}.}
  \label{new_gaia_36}
\end{figure*}

\clearpage
\section{inner halo known clusters}

\begin{figure*}[!h]
  \centering
  \includegraphics[width=0.75\textwidth]{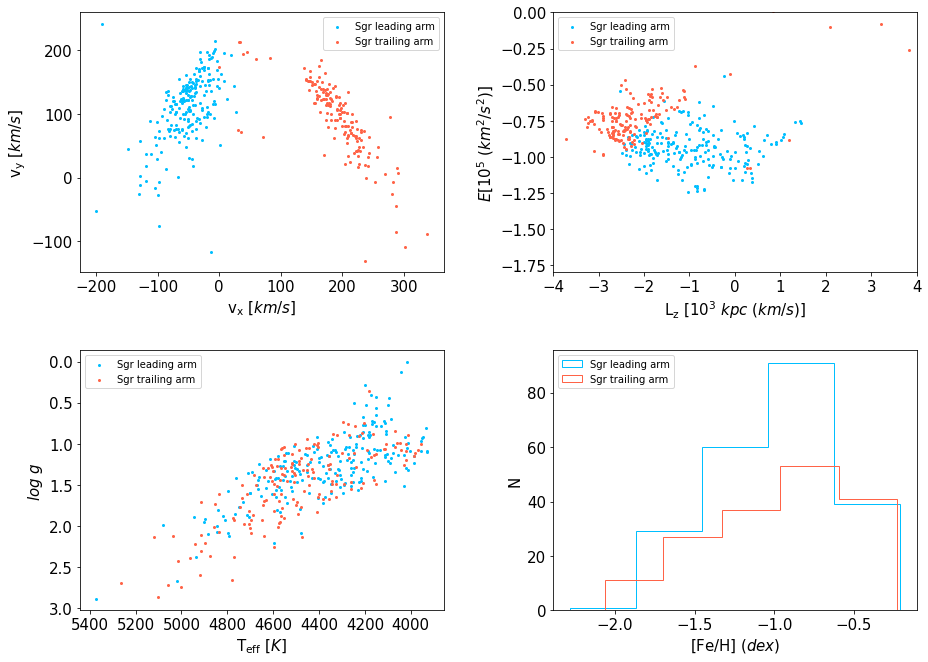}
  \caption{Detailed analysis of the Sgr obtained in our GS$^{3}$ Hunter crossed matched with the result of \citet{2022A&A...666A..64R}.The blue color indicates the leading arm and the red color indicates the trailing arm. Top left panel: The velocity distribution relative to the Galactic center. Top right panel: Distribution of the angular momentum and the energy (IOM) in the Sgr. Bottome left panel: the Kiel diagram of Sgr. Bottom right panel: we present the metallicity distribution of Sgr.}
  \label{Sgr_more}  
\end{figure*}

\begin{figure*}[!h]
  \centering
  \includegraphics[width=0.75\textwidth]{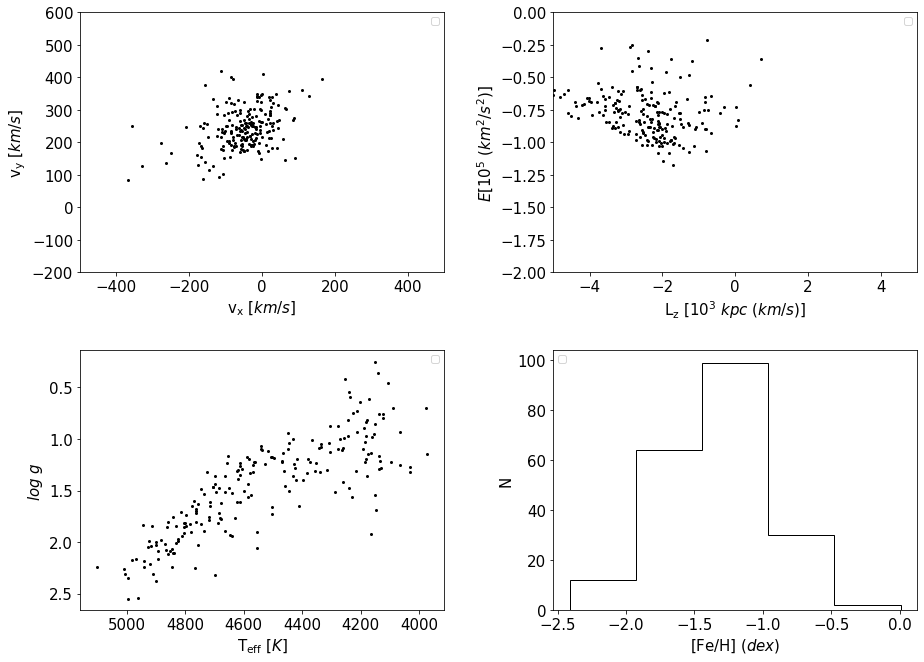}
  \caption{Detailed analysis of the virgo overdensity (VOD) obtained in our GS$^{3}$ Hunter. Top left panel: The velocity distribution relative to the Galactic center. Top right panel: Distribution of the angular momentum and the energy (IOM) in the VOD. Bottome left panel: the Kiel diagram of VOD. Bottom right panel: we present the metallicity distribution of VOD.}
  \label{VOD}  
\end{figure*}

\begin{figure*}[!h]
  \centering
  \includegraphics[width=0.75\textwidth]{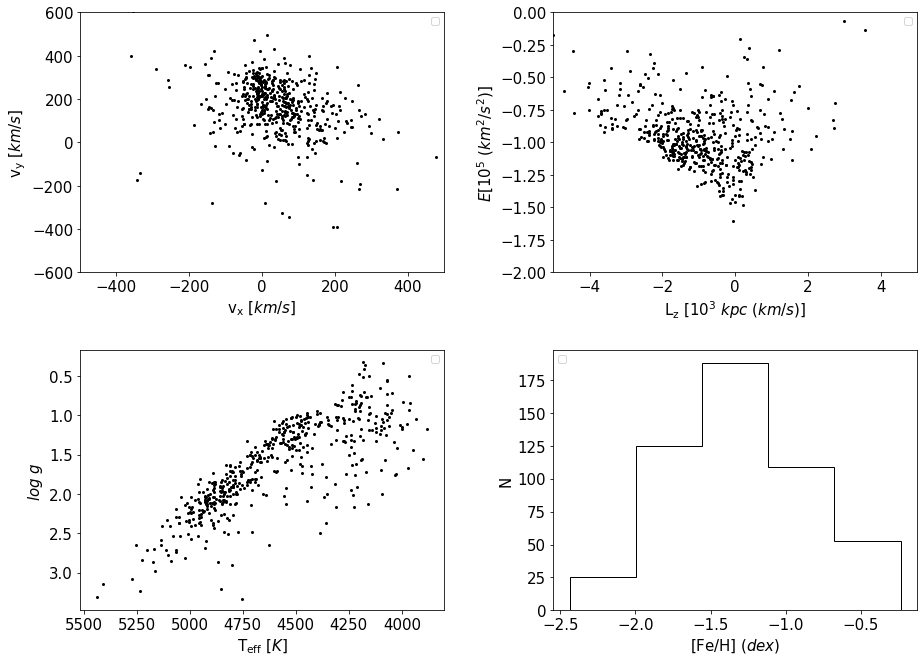}
  \caption{Detailed analysis of the Hercules-Aquila cloud (HAC) obtained in our GS$^{3}$ Hunter. Top left panel: The velocity distribution relative to the Galactic center. Top right panel: Distribution of the angular momentum and the energy in the HAC, which is more compact as shown. Bottom left panel: the Kiel diagram of HAC. Bottom right panel: we present the metallicity distribution of HAC.}
  \label{HAC}
\end{figure*}

\begin{figure*}[!h]
  \centering
  \includegraphics[width=0.9\textwidth]{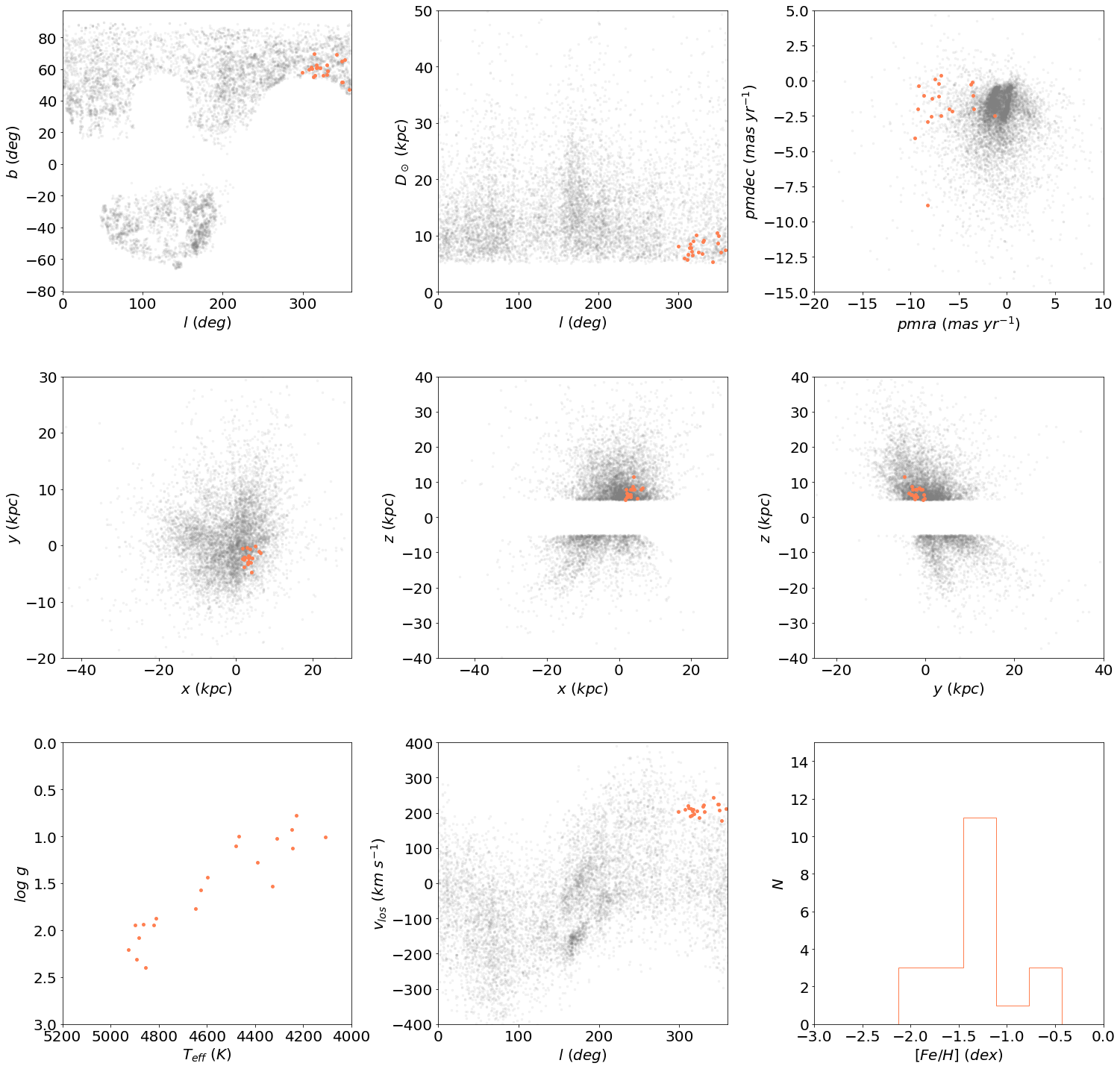}
  \caption{The first row shows the sky position distribution of cluster $\#$14 in our inner halo search, distance distribution with longitude, and the third one shows the proper motion of this structure. The second row shows the cluster $\#$14 distribution in 3D phase space. And the first panel of the third row shows the Kiel distribution, the second panel shows its line-of-sight velocity distribution with longitude, and the third panel shows the [Fe/H] distribution. For reference, the gray background shows all our inner halo samples. This structure cluster $\#$14 might be C-22 mentioned in \citet{2022MNRAS.516.5331M}, note that if we want to investigate the inner halo more carefully, we need to redesign the search strategy, we defer the more detailed analysis of more inner halo structures to a later contribution.}.
  \label{lamost_14}
\end{figure*}

\begin{figure*}[!h]
  \centering
  \includegraphics[width=0.9\textwidth]{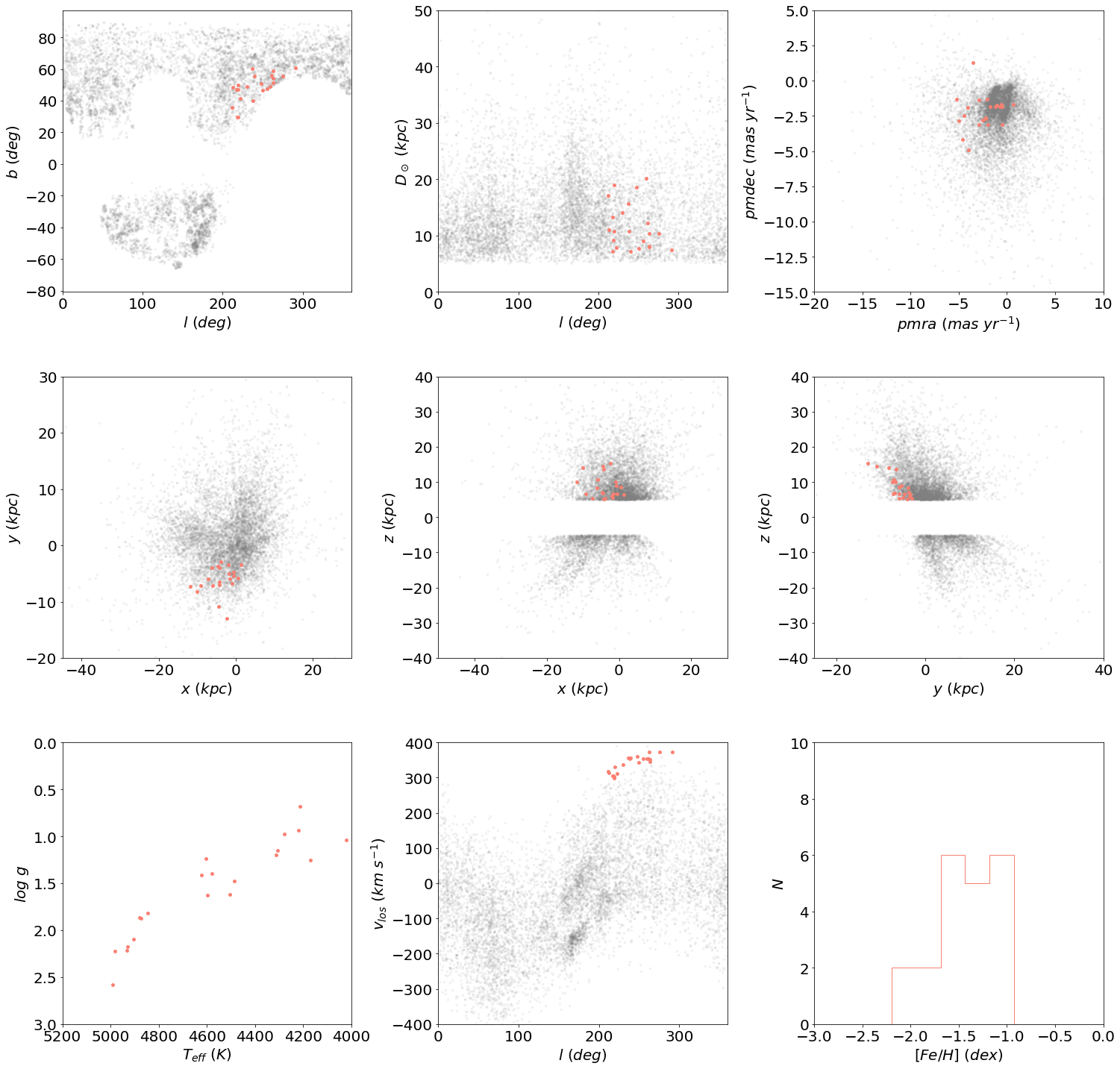}
 \caption{Cluster $\#$15 (inner halo quick search). Similar to the Figure \ref{lamost_14}. This structure might be the Slidr mentioned in \citet{2021ApJ...914..123I}.}
  \label{lamost_15}
\end{figure*}

\begin{figure*}[!h]
  \centering
  \includegraphics[width=0.9\textwidth]{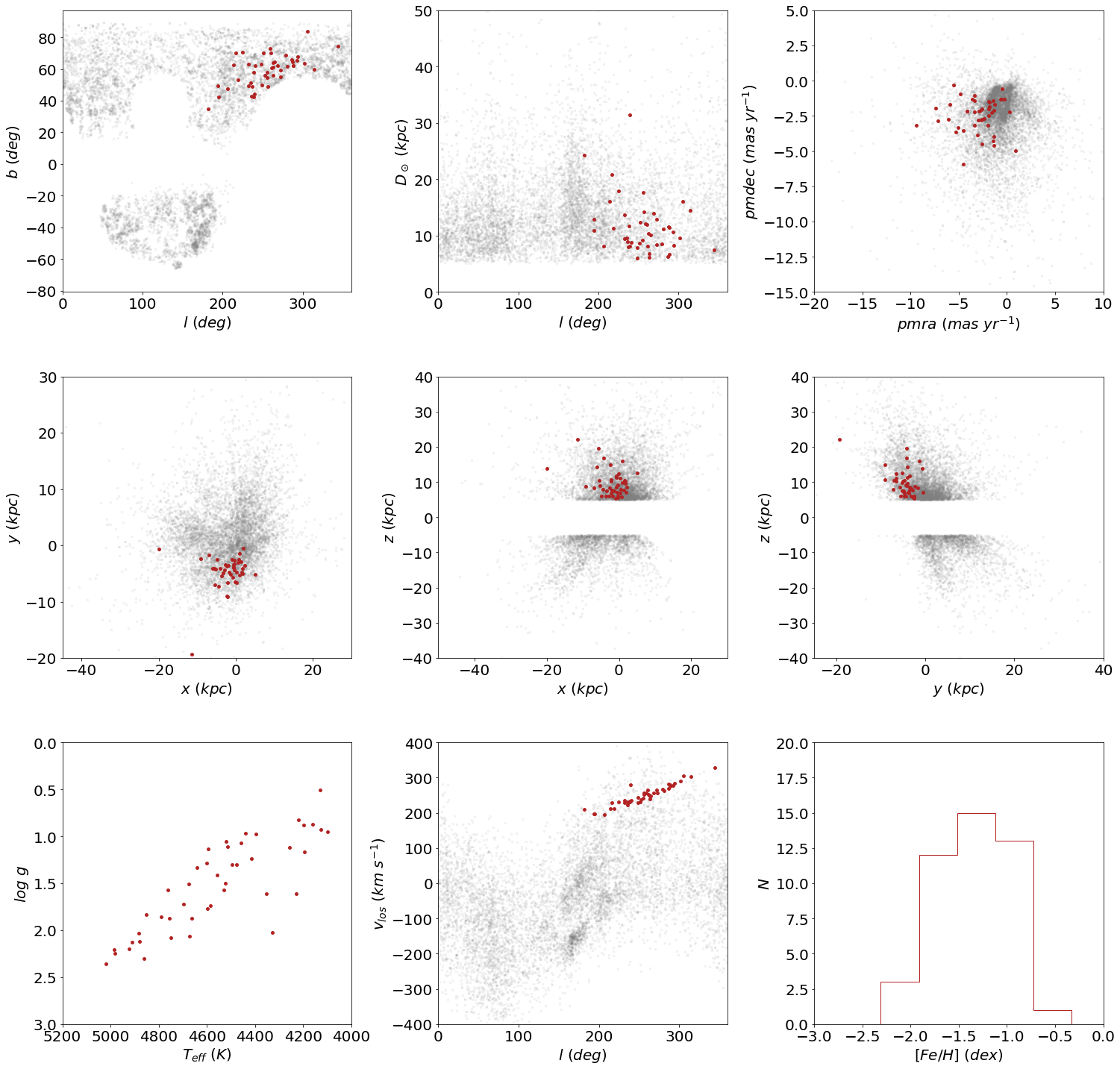}
  \caption{Cluster $\#$22 (inner halo). Similar to the Figure \ref{lamost_14}. This structure might be the combination of Slidr and Sylgr (see \citet{2021ApJ...914..123I}).}
  \label{lamost_22}
\end{figure*}

\begin{figure*}[!h]
  \centering
  \includegraphics[width=0.9\textwidth]{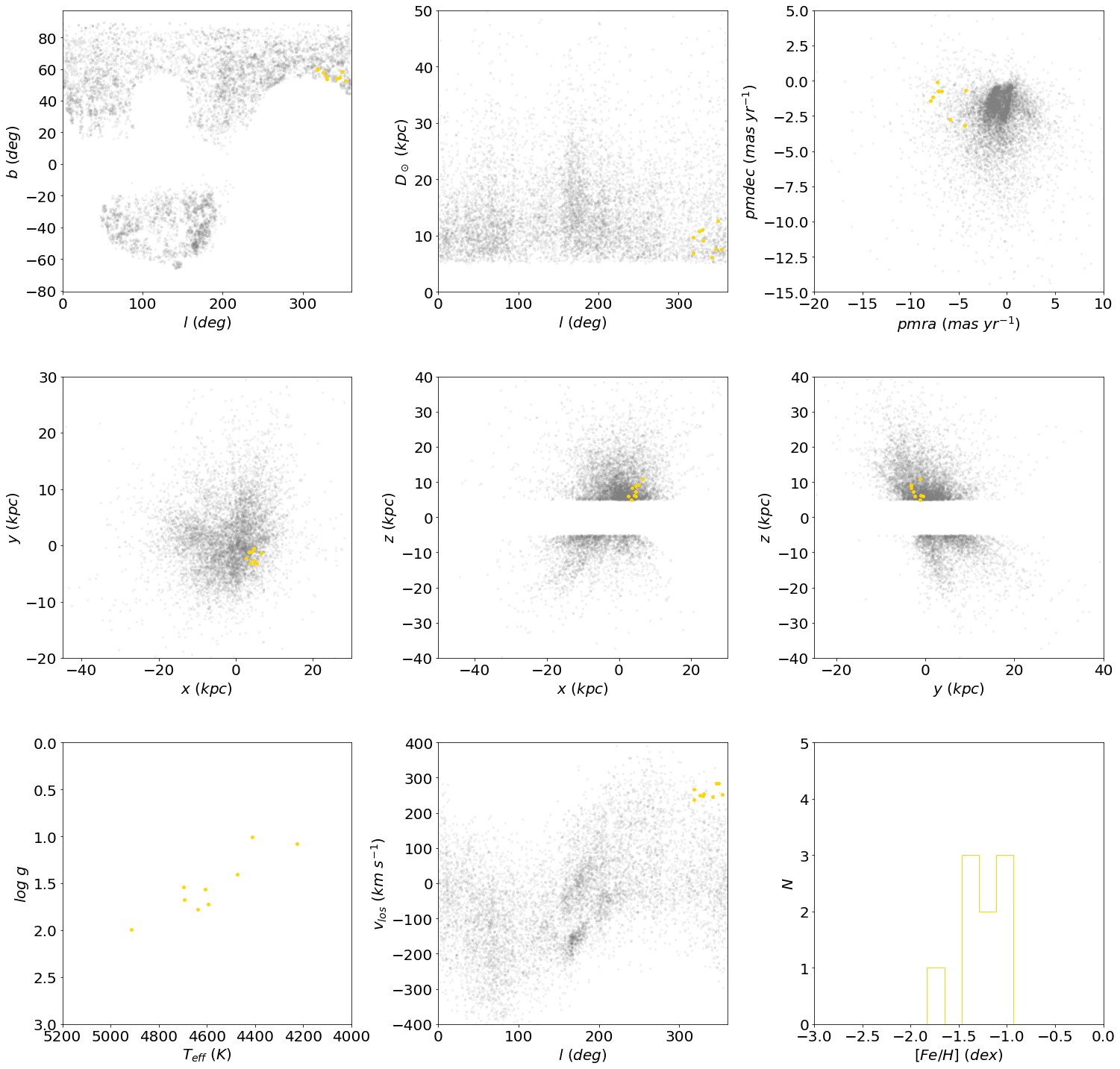}
  \caption{Cluster $\#$26 (inner halo). Similar to the Figure \ref{lamost_14}. This structure must be a part of Gaia-1 (see \citet{2022MNRAS.516.5331M}).}
  \label{lamost_26}
\end{figure*}

\begin{figure*}[!h]
  \centering
  \includegraphics[width=0.9\textwidth]{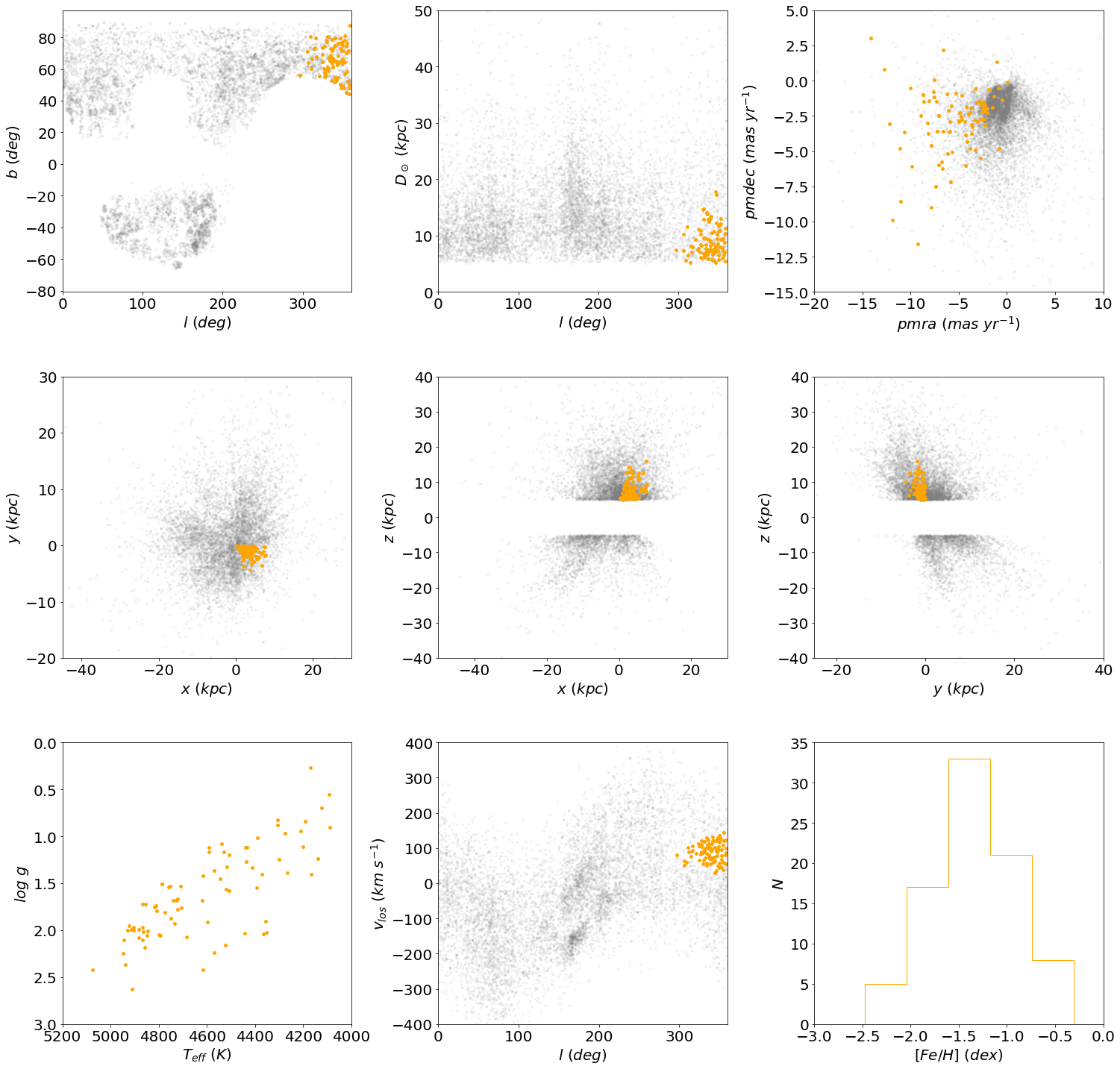}
  \caption{Cluster $\#$33 (inner halo). Similar to the Figure \ref{lamost_14}. This structure contains C-1 (see \citet{2022MNRAS.516.5331M}).}
  \label{lamost_33}
\end{figure*}

\begin{figure*}[!h]
  \centering
  \includegraphics[width=0.9\textwidth]{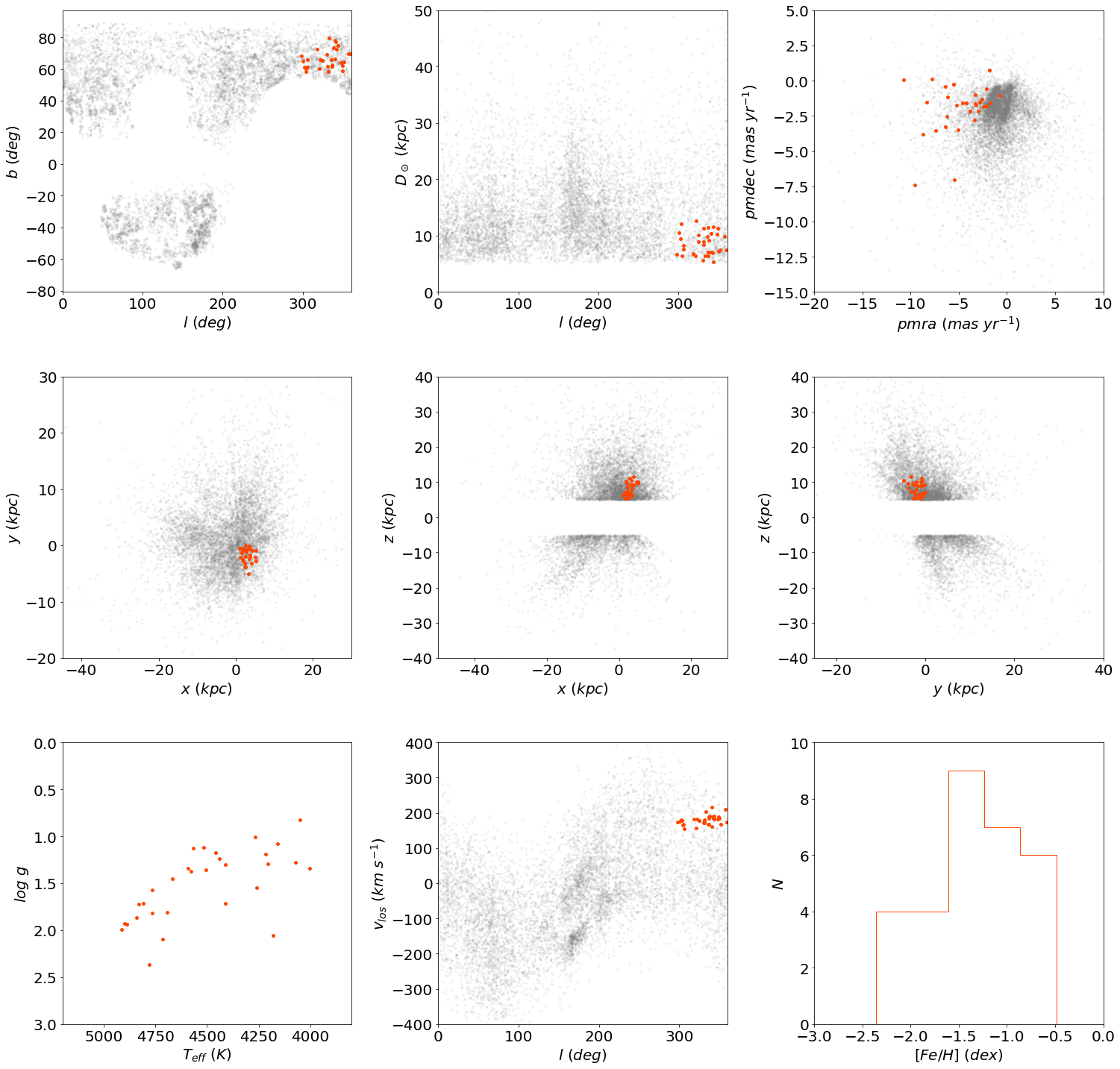}
  \caption{Cluster $\#$35 (inner halo). Similar to the Figure \ref{lamost_14}. This structure might be the combination of C-1 and Fjörm structures (see \citet{2021ApJ...914..123I}).}
  \label{lamost_35}
\end{figure*}

\clearpage
\section{Tables for substructures}
\begin{table*}[!t]%调节图片位置，h：浮动；t：顶部；b:底部；p：当前位置
\centering
    \caption{FIRE simulation validations for structures.}%title center
    \begin{tabular}{ccccc}%表格中的数据居中，c的个数为表格的列数
    \hline\hline\noalign{\smallskip}	
    N(cluster) & corresponding structure & fraction (of corresponding structure)\\
    \noalign{\smallskip}\hline\noalign{\smallskip}
    0 & 3 & 35.29\%\\
    1 & 9 & 52.03\%\\
    2 & 2 & 53.65\%\\
    3 & 0 & 49.83\%\\
    4 & 1 & 67.81\%\\
    5 & 6 & 43.12\%\\
    6 & 3 & 15.02\%\\
    7 & 3 & 15.65\%\\
    8 & 6 & 23.30\%\\
    % 9 & 7 & 2.64\%\\
    9 & 2 & 3.97\%\\		
    10 & 8 & 5.36\%\\
    % 12 & 5 & 9.52\%\\
    11 & 4 & 34.24\%\\
    12 & 9 & 17.12\%\\
    13 & 0 & 9.78\%\\
    14 & 3 & 7.59\%\\
    15 & 2 & 13.50\%\\
    16 & 0 & 2.87\%\\
    17 & 0 & 3.93\%\\
    18 & 3 & 3.18\%\\
    19 & 8 & 19.24\%\\
    20 & 0 & 3.70\%\\
    21 & 8 & 21.89\%\\
    22 & 0 & 1.16\%\\
    23 & 8 & 3.57\%\\
    24 & 8 & 7.21\%\\
    25 & 3 & 16.66\%\\
    26 & 1 & 14.42\%\\
    27 & 2 & 4.74\%\\
    28 & 2 & 1.52\%\\
    29 & 9 & 1.84\%\\
    30 & 8 & 0.63\%\\
    31 & 8 & 0.94\%\\
    32 & 6 & 32.71\%\\
    \noalign{\smallskip}\hline
    \label{FIRE}
    \end{tabular}
\end{table*}

\clearpage
\begin{table*}[!t]%调节图片位置，h：浮动；t：顶部；b:底部；p：当前位置
	\centering
	\caption{The definition of each clusters/groups (local halo)}%title center
	\label{tab:2} 
        % \begin{threeparttable}
    	\begin{tabular}{ccccccccccc}%表格中的数据居中，c的个数为表格的列数
    		\hline\hline\noalign{\smallskip}	
    		N & {\tt\emph x} (kpc) & {\tt\emph y} (kpc) & {\tt\emph z} (kpc) & {\tt\emph v$_x$} (km~s$^{-1}$) & {\tt\emph v$_y$} (km~s$^{-1}$)  & {\tt\emph v$_z$} (km~s$^{-1}$) & star fraction & Yes/No\\
    		\noalign{\smallskip}\hline\noalign{\smallskip}
    		0 & (-2.5, 2.4) & (-2.4, 2.3) & (-2.4, 2.5) & (-466.3, -112.4) & (-547.7, -125.4) & (-40.1, 146.7) & 75.4\% &  No\\
    		1 & (-2.4, 2.4) & (-2.3, 2.4) & (-2.4, 2.5) & (78.1, 442.2) & (-658.1, -96.8) & (-284.5, 10.5) & 83.7\% & No \\
    		2 & (-2.5, 2.5) & (-2.5, 2.4) & (-2.4, 2.5) & (-369.8, -67.0) & (-609.2, -94.4) & (-22.2, 337.7) & 56.1\% & No \\
    		3 & (-2.5, 2.5) & (-2.5, 2.4) & (-2.5, 2.5) & (35.0, 286.1) & (-583.1, -156.6) & (-26.3, 173.7) & 92.6\% & No \\
    		4 & (-2.4, 2.4) & (-2.4, 2.4) & (-2.4, 2.5) & (-235.9, -24.8) & (-737.1, -135.9) & (-83.2, 417.7) & 63.9\% & No \\
    		5 & (-2.4, 2.4) & (-2.4, 2.2) & (-2.3, 2.5) & (-415.1, -67.8) & (-605.9, -121.9) & (-183.4, 24.8) & 65.1\% & No \\
    		6 & (-2.3, 2.2) & (-2.4, 2.3) & (-2.4, 2.4) & (-140.1, 6.5) & (-721.3, -206.9) & (-141.8, -2.7) & 97.0\% & Yes \\
    		7 & (-2.2, 2.4) & (-2.2, 2.3) & (-2.4, 2.4) & (2.8, 444.9) & (-548.1, -0.6) & (48.3, 424.6) & 75.0\% & No \\
    		8 & (-2.0, 2.2) & (-2.2, 1.9) & (-1.9, 2.3) & (113.6, 350.8) & (-267.9, -49.8) & (-270.6, -79.9) & 96.5\% & Yes \\
    		9 & (-2.3, 2.3) & (-2.4, 2.3) & (-2.4, 2.5) & (-114.7, 145.5) & (-726.2, -177.4) & (-0.5, 300.5) & 98.2\% & Yes \\
    		10 & (-2.3, 2.3) & (-2.3, 2.3) & (-2.4, 2.5) & (96.2, 463.8) & (-538.3, -13.9) & (-22.9, 134.4) & 80.6\% & No \\		
    		11 & (-1.6, 2.3) & (-1.8, 1.5) & (-1.3, 2.3) & (28.3, 450.9) & (-108.5, 222.4) & (-232.9, -4.7) & 96.3\% & Yes \\
    		12 & (-2.4, 2.4) & (-2.4, 2.5) & (-2.4, 2.4) & (-95.7, 124.7) & (-763.7, -152.5) & (-195.2, 408.9) & 82.5\% & No \\
    		13 & (-2.4, 2.3) & (-2.2, 2.4) & (-1.9, 2.2) & (-531.1, -151.9) & (-249.0, 126.9) & (-234.6, -0.9) & 95.2\% & Yes \\
    		14 & (-1.8, 2.3) & (-2.1, 1.9) & (-1.8, 2.4) & (-384.1, 148.9) & (-342.5, 33.7) & (97.1, 515.8) & 96.1\% & Yes \\
    		15 & (-2.5, 2.4) & (-1.7, 2.4) & (-1.7, 2.4) & (-526.1, -14.9) & (-133.9, 197.6) & (-36.4, 317.2) & 96.3\% & Yes \\
    		16 & (-1.8, 2.4) & (-2.5, 2.3) & (-2.0, 2.5) & (39.0, 243.4) & (-439.6, -119.9) & (-292.4, -70.4) & 98.9\% & Yes \\
    		17 & (-1.8, 2.4) & (-2.4, 2.3) & (-2.1, 2.5) & (-376.9, 96.3) & (-382.2, 92.9) & (-524.9, -93.9) & 95.6\% & Yes \\
    		18 & (-1.2, 2.4) & (-1.6, 1.6) & (-1.5, 2.2) & (138.8, 392.9) & (-95.7, 6.6) & (-222.8, -194.) & 98.9\% & Yes \\
    		19 & (-2.3, 2.5) & (-2.2, 2.3) & (-2.3, 2.4) & (119.1, 387.2) & (-377.8, -82.3) & (-145.1, 0.4) & 76.2\% & No \\
    		20 & (-2.5, 2.4) & (-2.4, 1.0) & (-0.5, 2.1) & (160.3, 495.7) & (-12.8, 79.0) & (-44.6, 33.6) & 97.2\% & Yes \\
    		21 & (-2.0, 2.3) & (-2.4, 2.3) & (-2.1, 2.2) & (-171.9, 57.1) & (-600.1, -62.4) & (-443.7, -98.9) & 85.8\% & No \\
    		22 & (-2.4, 2.5) & (-2.0, 2.4) & (-2.3, 2.4) & (-570.2, -154.0) & (-43.3, 10.4) & (-70.2, 217.5) & 83.4\% & No \\
    		23 & (-2.4, 2.3) & (-2.3, 2.2) & (-2.2, 2.4) & (-54.7, 299.9) & (-616.3, -98.5) & (-442.3, -73.5) & 98.4\% & Yes \\
    		24 & (-1.7, 2.0) & (-1.6, 2.2) & (-1.9, 2.5) & (-249.6, 296.1) & (-154.5, 271.2) & (-460.9, 476.9) & 80.8\% & No \\
    		25 & (-2.4, 2.4) & (-2.4, 2.2) & (-2.0, 2.3) & (132.1, 493.5) & (-235.1, -9.7) & (-170.1, 18.7) & 98.4\% & Yes \\
    		26 & (-2.3, 2.4) & (-2.3, 2.3) & (-2.2, 2.5) & (48.7, 368.5) & (-583.7, -110.8) & (-242.6, -52.3) & 92.1\% & No \\
    		27 & (-2.3, 2.5) & (-2.2, 2.2) & (-2.3, 2.4) & (-212.8, 63.5) & (-702.4, -189.6) & (-219.4, -44.5) & 97.8\% & Yes \\
    		28 & (-2.2, 2.4) & (-2.4, 2.3) & (-2.4, 2.4) & (-440.8, -9.7) & (-407.9, -41.9) & (-385.4, -33.3) & 67.8\% & No \\
    		29 & (-2.2, 2.5) & (-2.1, 2.3) & (-2.2, 2.4) & (-282.7, 386.6) & (-411.2, 82.8) & (-394.4, 108.9) & 87.8\% & No \\
    		30 & (-2.4, 2.5) & (-2.5, 2.4) & (-2.4, 2.4) & (-11.8, 254.3) & (-718.9, -119.2) & (-225.9, 209.3) & 97.5\% & Yes \\
    		31 & (-2.2, 2.3) & (-2.0, 2.3) & (-1.9, 2.4) & (-15.5, 497.8) & (-400.4, 4.6) & (-6.7, 402.9) & 97.5\% & Yes \\
    		32 & (-2.2, 2.4) & (-2.3, 2.3) & (-2.3, 2.4) & (-320.1, 96.1) & (-533.4, -91.9) & (-505.8, -55.9) & 90.5\% & No \\
    		33 & (-2.1, 2.4) & (-2.1, 2.3) & (-2.2, 2.4) & (111.0, 396.8) & (-345.0, -75.7) & (-10.9, 127.6) & 91.3\% & No \\
    		34 & (-2.5, 2.2) & (-2.4, 2.3) & (-2.5, 2.4) & (-69.1, 19.9) & (-778.3, -211.2) & (-81.7, 71.2) & 98.4\% & Yes \\
    		35 & (-2.3, 1.8) & (-1.6, 1.8) & (-1.9, 2.3) & (14.2, 388.8) & (-128.2, 148.9) & (-432.7, 41.4) & 97.8\% & Yes \\
    		36 & (-1.0, 1.8) & (-1.1, 2.0) & (-2.2, 1.4) & (-118.1, 341.3) & (56.3, 259.0) & (-371.2, 247.1) & 97.4\% & Yes\\
    		37 & (-2.4, 2.4) & (-2.4, 2.3) & (-2.3, 2.4) & (226.9, 483.9) & (-444.4,-59.1) & (-117.5, 81.7) & 99.8\% & Yes \\
    		\noalign{\smallskip}\hline
    	\end{tabular}
            \begin{tablenotes}
            \footnotesize              
                \item{[1] first row N shows the sequence number of the group/cluster in our local halo results. The second to seventh rows we show the range of {\tt\emph x}, {\tt\emph y}, {\tt\emph z}, {\tt\emph v$_x$}, {\tt\emph v$_y$}, {\tt\emph v$_z$} for each group/cluster separately. The eighth row we show the star fraction of each group/cluster define a group candidate according to the 95\% criterion. The last  column is the whether or not the cluster is a candidate.}   
            \end{tablenotes}           
        % \end{threeparttable}
        \label{tab:2}
\end{table*}

\clearpage
\begin{table*}[!t]%调节图片位置，h：浮动；t：顶部；b:底部；p：当前位置
	\centering
	\caption{The definition of each clusters/groups (inner halo)}%title center
	\label{tab:3}  
	\begin{tabular}{ccccccccc}%表格中的数据居中，c的个数为表格的列数
		\hline\hline\noalign{\smallskip}	
		N & {\tt\emph l} (deg) & {\tt\emph b} (deg) & {\tt\emph d} (kpc) & {\tt\emph r$_v$} (km~s$^{-1}$) & {\tt\emph v$_l$} (km~s$^{-1}$)  & {\tt\emph v$_b$} (km~s$^{-1}$) & star fraction & Yes/No\\
		\noalign{\smallskip}\hline\noalign{\smallskip}
		0 & (20.2, 337.7) & (-65.2, 89.4) & (5.0, 51.8) & (-249.0, 97.3) & (-12.2, 39.3) & (-9.9, 11.7) & 73.9\% &  No\\
		1 & (68.2, 240.7) & (-65.6, 85.4) & (5.3, 44.5) & (-153.9, 19.0) & (-4.5, 32.0) & (-8.0, 14.4) & 92.9\% & No \\
		2 & (35.8, 359.4) & (-48.1, 89.2) & (5.0, 69.8) & (-296.8, 77.1) & (-10.1, 17.1) & (-10.1, 15.1) & 86.9\% & No \\
		3 & (171.3, 359.9) & (26.7, 88.6) & (5.1, 60.5) & (-124.6, 157.8) & (-58.6, 58.8) & (-60.7, 12.2) & 98.2\% & Yes \\
		4 & (1.2, 133.8) & (-58.8, 86.4) & (5.5, 49.8) & (-335.9, -141.8) & (-11.5, 5.4) & (-13.1, 18.5) & 95.2\% & Yes \\
		5 & (21.8, 358.8) & (-16.7, 88.7) & (5.3, 58.0) & (-217.1, 177.5) & (-8.4, 17.2) & (-10.1.4, 22.9) & 81.5\% & No \\
		6 & (38.2, 355.2) & (20.7, 89.1) & (5.1, 34.9) & (21.4, 209.3) & (-8.6, 9.2) & (-10.1, 12.4) & 90.4\% & No \\
		7 & (5.5, 162.9) & (-53.1, 82.3) & (6.3, 58.7) & (-479.4, -276.0) & (-12.4, 10.8) & (-22.1, 14.0) & 99.2\% & Yes \\
		8 & (0.1, 169.5) & (-65.3, 88.4) & (5.4, 53.7) & (-310.9, -76.8) & (-12.4, 15.6) & (-11.3, 8.5) & 92.2\% & No \\
		9 & (3.6, 187.5) & (-61.4, 82.6) & (5.5, 36.8) & (-278.9, -20.7) & (-8.3, 5.9) & (-6.9, 15.9) & 81.8\% & No \\
		10 & (0.2, 182.9) & (-46.7, 87.5) & (5.1, 36.7) & (38.6, 333.1) & (-16.0, 5.9) & (-30.7, 9.2) & 95.1\% & Yes \\		
		11 & (50.3, 256.8) & (-60.2, 63.7) & (6.7, 118.1) & (17.3, 251.7) & (-2.8, 12.2) & (-9.4, 9.1) & 89.3\% & No \\
		12 & (67.1, 359.1) & (-58.3, 55.3) & (6.6, 63.8) & (-43.4, 89.6) & (-2.0, 25.9) & (-54.3, 8.4) & 99.4\% & Yes \\
		13 & (50.3, 173.8) & (-61.7, 23.3) & (5.8, 45.6) & (-281.8, -84.6) & (-6.9, 7.4) & (-10.4, 1.8) & 98.9\% & Yes \\
		14 & (299.2, 358.1) & (47.0, 69.9) & (5.3, 10.5) & (177.3, 243.0) & (-18.9, -2.2) & (-2.5, 12.2) & 100.0\% & Yes \\
		15 & (211.9, 291.0) & (29.4, 60.5) & (7.1, 20.1) & (298.9, 373.2) & (-3.7, 2.8) & (-6.3, 0.6) & 100.0\% & Yes \\
		16 & (0.1, 174.8) & (15.9, 85.2) & (5.2, 44.7) & (-241.2, 10.9) & (-17.8, 3.8) & (-11.5, 8.4) & 70.1\% & No \\
		17 & (157.7, 358.5) & (6.3, 63.2) & (6.1, 57.8) & (28.2, 120.4) & (-4.3, 22.5) & (-14.3, 0.5) & 93.1\% & No \\
		18 & (0.4, 187.4) & (-63.4, 86.3) & (5.4, 40.3) & (-272.6, -1.5) & (-16.1, 10.1) & (-11.8, 8.5) & 70.6\% & No \\
		19 & (214.1, 301.3) & (39.6, 76.8) & (5.5, 18.2) & (231.1, 343.6) & (-9.5, 2.1) & (-7.9, -1.4) & 97.2\% & Yes \\
		20 & (0.7, 160.4) & (33.3, 85.9) & (5.4, 37.9) & (-292.4, -54.2) & (-9.5, 1.6) & (-3.7, 18.6) & 98.1\% & Yes \\
		21 & (155.9, 331.1) & (40.2, 87.1) & (6.1, 25.2) & (132.3, 249.7) & (-7.0, -2.5) & (-8.7, -0.6) & 97.1\% & Yes \\
		22 & (181.9, 344.6) & (34.7, 83.7) & (6.0, 31.5) & (194.1, 328.8) & (-7.7, 3.8) & (-7.4, -0.1) & 97.8\% & Yes \\
		23 & (161.1, 338.6) & (11.7, 83.8) & (5.3, 55.8) & (91.8, 206.6) & (-7.1, 8.3) & (-18.2, 0.3) & 98.7\% & Yes \\
		24 & (1.1, 223.2) & (24.4, 88.8) & (5.0, 51.9) & (-244.3, 66.6) & (-14.7, 10.9) & (-7.7, 18.2) & 79.6\% & No \\
		25 & (170.6, 278.9) & (21.8, 84.3) & (6.3, 23.1) & (181.3, 304.7) & (-1.5, 8.6) & (-7.6, -0.1) & 97.9\% & Yes \\
		26 & (318.4, 353.7) & (52.7, 60.2) & (6.1, 12.6) & (236.6, 284.7) & (-8.0, -3.7) & (0.1, 3.9) & 100\% & Yes \\
		27 & (188.8, 329.9) & (26.7, 82.8) & (5.5, 18.6) & (175.0, 317.5) & (-15.7, 3.2) & (-4.9, 0.2) & 98.3\% & Yes \\
		28 & (0.0, 160.6) & (23.3, 88.4) & (5.1, 58.4) & (-265.2, -43.5) & (-13.1, 6.4) & (-4.2, 14.2) & 95.1\% & Yes \\
		29 & (1.4, 137.3) & (-57.5, 87.1) & (5.1, 34.3) & (-302.6, -204.1) & (-16.1, 8.4) & (-8.2, 11.1) & 96.0\% & Yes \\
		30 & (0.1, 166.1) & (-61.6, 83.1) & (5.5, 66.9) & (-259.9, -21.7) & (-14.6, 12.4) & (-10.0, 18.2) & 82.3\% & No \\
		31 & (0.1, 172.5) & (16.9, 88.6) & (5.1, 48.4) & (-227.0, 36.5) & (-13.9, 11.2) & (-8.5, 12.4) & 86.9\% & No \\
		32 & (131.6, 355.8) & (21.9, 84.7) & (5.7, 40.9) & (124.2, 312.2) & (-5.2, 7.5) & (-9.5, 3.9) & 98.2\% & Yes \\
		33 & (296.9, 358.8) & (44.1, 87.7) & (5.1, 17.7) & (30.6, 144.3) & (-23.5, -0.1) & (-7.3, 11.6) & 98.9\% & Yes \\
		34 & (-277.0, 349.6) & (54.1, 74.6) & (6.8, 20.1) & (221.3, 358.1) & (-5.1, -0.8) & (-7.8, 2.7) & 100.0\% & Yes \\
		35 & (298.1, 359.1) & (58.2, 79.6) & (5.3, 12.5) & (155.7, 215.9) & (-9.7, -1.2) & (-7.2, 5.9) & 97.1\% & Yes \\
		36 & (160.5, 359.8) & (27.9, 78.2) & (5.3, 73.9) & (-41.8, 263.1) & (-14.1, 3.8) & (-8.4, 7.7) & 96.6\% & Yes\\
		37 & (199.8, 358.5) & (20.5, 56.3) & (-7.4, 20.5) & (101.3, 210.3) & (-8.3, 3.3) & (-2.7, 5.6) & 72.4\% & No \\
		38 & (47.8, 173.4) & (-66.0, -14.4) & (6.8, 40.1) & (-213.9, -37.1) & (-5.1, 12.3) & (-15.8, 0.2) & 96.2\% & Yes \\
		39 & (224.6, 332.9) & (41.6, 76.4) & (6.1, 24.4) & (170.2, 265.7) & (-7.3, 1.5) & (-7.2, 1.6) & 97.8\% & Yes \\
		40 & (50.3, 199.4) & (-66.1, -17.6) & (5.9, 42.1) & (-236.1, -17.5) & (-2.9, 8.0) & (-10.7, -0.1) & 93.8\% & No \\
		41 & (139.1, 299.3) & (23.9, 86.4) & (6.0, 24.9) & (173.8, 344.6) & (-3.7, 12.8) & (-11.2, 1.5) & 98.3\% & Yes \\
		42 & (182.4, 359.5) & (26.9, 85.5) & (5.1, 24.5) & (128.3, 390.9) & (-7.8, 4.6) & (-6.2, 5.2) & 91.7\% & No \\
		43 & (259.3, 260.9) & (49.2, 58.5) & (7.8, 17.8) & (390.9, 433.4) & (-1.2, 2.4) & (-8.9, -6.5) & 100.0\% & Yes \\
		44 & (55.1, 185.9) & (-64.2, -15.8) & (5.9, 58.7) & (-176.9, 0.46) & (-9.1, 21.8) & (-11.7, 0.7) & 96.9\% & Yes \\
		\noalign{\smallskip}\hline
        \end{tabular}
        \begin{tablenotes}
        \footnotesize              
            \item{[1] first column N shows the sequence number of the group/cluster in our inner halo results. The second to seventh columns we show the range of {\tt\emph l}, {\tt\emph b}, {\tt\emph d}, {\tt\emph r$_v$}, {\tt\emph v$_l$}, {\tt\emph v$_b$} for each group/cluster separately. The eighth column we show the star fraction of each group/cluster to define the candidate of structures according to the 95\% criterion. The last column shows whether or not the group is a candiadate.}    
        \end{tablenotes}
	\label{tab:3}
\end{table*}

\clearpage

\end{document}